\documentclass{aa}
\usepackage{graphicx}
\usepackage{txfonts}
\usepackage[english]{babel}
\usepackage{amsmath}
\usepackage[normalem]{ulem}
\usepackage[breaklinks]{hyperref}
\usepackage[textwidth=1.8cm, textsize=small]{todonotes}
\usepackage[math, probs, units, fig, names, todonotes, prettylinks]{mf}

\providecommand{\teff}{\ensuremath{\textrm{T}_{\textrm{eff}}}\xspace}
\providecommand{\feh}{\ensuremath{\textrm{[Fe/H]}}\xspace}
\providecommand{\logg}{\ensuremath{\log\,\textrm{g}}\xspace}

\providecommand{\Rzero}{\ensuremath{\textrm{R}_{0}}\xspace}
\providecommand{\Azero}{\ensuremath{\textrm{A}_{0}}\xspace}
\providecommand{\dmod}{\ensuremath{\mu\xspace}}
\providecommand{\mags}{\ensuremath{\,\textrm{mag}\xspace}}

\providecommand{\dif}{\ensuremath{\textrm{d}}}

\providecommand{\pars}{\ensuremath{\boldsymbol{\theta}}}
\providecommand{\gmag}{\ensuremath{\textrm{G}}}
\providecommand{\bpmag}{\ensuremath{\textrm{BP}}}

\providecommand{\jitter}{\ensuremath{\eta}}

\providecommand{\given}{\ensuremath{\hspace{0.05em}\mid\hspace{0.05em}}}

\providecommand{\gdr}[1]{Gaia~DR{#1}}

\providecommand{\gmag}{\ensuremath{G}}

\providecommand{\parallax}{\ensuremath{\varpi}}
\providecommand{\sigparallax}{\ensuremath{\sigma_{\varpi}}}

\providecommand{\gdr}[1]{Gaia~DR{#1}}
\def\apogee{APOGEE~DR14}
\def\ges{Gaia-ESO~iDR5}
\def\galah{GALAH~DR2}
\def\lamost{LAMOST~DR4}
\def\rave{RAVE~DR5}
\def\sdss{SDSS~DR14}
\def\kicA{Kepler raw catalog}
\def\kicB{Kepler Input Catalog}
\def\redclump{APOGEE red clump catalog}
\def\cu6{CU6 Auxiliary catalog}
\def\obRamirez{OB stars \citep{RamirezAgudelo2017}}
\def\obSimon{OB stars \citep{SimonDiaz2017}}
\def\ntotvstcat{6\,292\,410}
\def\ntotcat{123\,097\,070}

\RequirePackage[normalem]{ulem}
\RequirePackage{color}
\definecolor{RED}{rgb}{1,0,0}
\definecolor{BLUE}{rgb}{0,0,1}

\begin{document}

   \title{Astrophysical Parameters from Gaia DR2, 2MASS \& AllWISE
   }
   \author{M. Fouesneau\inst{1}
          \and R. Andrae\inst{1}
          \and T. Dharmawardena \inst{1}
          \and J. Rybizki\inst{1}
          \and C. A. L. Bailer-Jones\inst{1}
          \and M. Demleitner\inst{2}
          }

   \institute{Max-Planck-Institut f\"ur Astronomie, K\"onigstuhl 17, 69117 Heidelberg, Germany --
              \email{fouesneau@mpia.de}
    \and
    Astronomisches Rechen-Institut, Zentrum f\"ur Astronomie der Universit\"at Heidelberg, Germany}

   \date{Received --; accepted --}


  \abstract
{Stellar physical and dynamical properties are essential knowledge to understanding the structure, formation, and evolution of our Galaxy.}
{We produced an all-sky uniformly derived catalog of stellar astrophysical parameters (APs; age, mass, temperature, bolometric luminosity, distance, dust extinction) to give insight into the physical properties of Milky-Way stars.}
{Exploiting the power of multi-wavelength and multi-survey observations from Gaia DR2 parallaxes and integrated photometry along with 2MASS and AllWISE photometry, we introduce an all-sky uniformly derived catalog of stellar astrophysical parameters, including dust extinction (\Azero) average grain size (\Rzero) along the line of sight,  for \ntotcat\ stars. In contrast with previous works, we do not use a Galactic model as prior in our analysis.
}
{We validate our results against other literature (e.g., benchmark stars, interferometry, Bayestar, StarHorse). The limited optical information in the Gaia photometric bands or the lack of ultraviolet or spectroscopic information renders the chemistry inference prior dominated.
We demonstrate that Gaia parallaxes bring sufficient leverage to explore the detailed structures of the interstellar medium in our Milky Way.}
{
In Gaia DR3, we will obtain the dispersed optical light information to break through some limitations of this analysis, allowing us to infer stellar chemistry in particular.  Gaia promises us data to construct the most detailed view of the chemo-dynamics of field star populations in our Galaxy.\\

Our catalog is available from the German Astrophysical Virtual Observatory (GAVO) at
\url{http://dc.g-vo.org/tableinfo/gdr2ap.main} where one can query it via ADQL and TAP (and soon in the Gaia Archive and CDS VizieR).
}

   \maketitle
%
\section{Introduction}\label{ref:intro}

Understanding the structure, formation, and evolution of our Galaxy requires studying in detail its stars both from the point of view of their dynamics and their physical properties.
The Gaia satellite's main technological advance is the accurate determination of parallaxes and proper motions for over one billion stars \citep{GaiaMission}. But without knowledge of the stellar properties, the resulting three-dimensional maps and velocity distributions one can derive from these are of limited value to understand our Galaxy. With kinematics and stellar properties, one can find members of streams and stellar populations to model the formation of Galactic disk \citep[e.g.][]{1998MNRAS.295..319M, 2013A&ARv..21...61R, 2019ApJ...872..152I}. Gaia is therefore equipped with two low-resolution spectrophotometers (BP/RP) and a high-resolution spectrograph (RVS). These instruments operate over the entire optical range and from 845–872 nm, respectively (see Gaia Collaboration, 2016 for the payload description).

The second Gaia data release (hereafter GDR2; Gaia Collaboration 2018b) contained 1.33 billion sources with positions, parallaxes, proper motions, and G-band photometry based on $22$ months of mission observations. \gdr{2} also includes the integrated G, BP, and RP fluxes.
\citet{GDR2Andrae2018} used just data from Gaia DR2 to infer stellar parameters. They provided the first inference of effective temperatures, radii, and luminosities for 160 million stars and line-of-sight extinctions for 88 million of them. This exercise demonstrated that three broad Gaia passbands (G, BP, RP) contain relatively little information to discriminate between effective temperature and interstellar extinction (see the referenced paper for details).

Deriving astrophysical properties (APs) of individual stars from their integrated photometry is a challenging project. {A single set of multi-wavelength photometry only contains limited amounts of information: whether limited by signal-to-noise, or physical stellar degeneracies, one needs additional measurements and assumptions to distinguish details of apparently similar stars.}
Previous work and catalogs of APs of Gaia sources are available, e.g., \citet{2017MNRAS.471..770M}, \citet{2017AJ....154..259S},  \citet{2017A&A...604A.108M,2018arXiv180406578M}, Bayestar \citep{Green2019}, StarHorse \citep{Anders2017}. The growing number of these catalogs originates from the difficulty of identifying the perfect set of assumptions.
Some may rely on open questions: what is the 3D geometry of the Milky Way? Where is the Galactic Bar? Although the catalogs globally agree, they differ in the details due to diverse and subtle systematic errors. For instance, \citet{BailerJones2018} discussed the significant effects of various prior distributions on estimating distances from parallaxes. But of course, these effects are more complex when inferring the temperature, gravity, chemical pattern, and other APs simultaneously, as these estimates are conditionally dependent on one another and the distance to that star.

The more stringent prior assumptions in an analysis, the greater the tensions between the models and the data {\citep[i.e., model discrepancy since models are not perfect representations of the real stars; e.g.][]{Kennedy2001}}\footnote{{One could make the parallel with the assumption of independent measurements from a single instrument: in very high signal-to-noise data, the systematics in the calibration challenge this assumption.}} Strains require compromises, which often lead to incorrect or incomplete results (e.g., significant uncertainties or biases). When results replicate our priors, one will wonder if we learn anything from the data. It is often hard to recognize false results in Gaia scale size catalogs (millions to billions of sources).
Our approach is to relax the commonly used strong assumptions in this present work. We will not assume the Milky-Way 3D stellar population structures, for instance. We analyze the stars for which \gdr{2} provides us with Gaia integrated photometry \citep{GDR2Riello2018, GDR2Evans2018} and where all external photometry is available from 2MASS \citep{2MASSsurvey, 2MASSCat} and AllWISE \citep{AllWiseCat}. Where available, we also make use of Gaia DR2 parallaxes \citep{GDR2Lindegren2018}. This represents about 120 million stars. We first describe the construction of the photometric input catalog in \autoref{sec:datasets}. In \autoref{sec:models}, we detail the modeling of the spectral energy distribution of stars (SED) and the
fitting procedure we use to infer stellar parameters (age, mass, temperature, luminosity, gravity, distance, dust extinction). In \autoref{sec:validation} we validate our results against other published results (e.g, benchmark stars, interferometry) and discuss our results in \autoref{sec:discussion} before  we
summarize in \autoref{sec:summary}.

\section{Cross-matched Photometric Catalog}\label{sec:datasets}

In the context of determining the astrophysical parameters (APs) of stars, one can achieve more reliable estimates by combining data from various surveys than those from an individual one.
One could combine multiple photometric, spectroscopic surveys or both \citep[e.g.,][]{Serenelli2013, Wang2016, Santiago2016, McMillan2018}. Stellar parameter estimates will then be more accurate and precise than those derived from any individual survey itself as long as the models are consistent with the data. However, one must be careful that systematics may arise if the data are not fully compatible: e.g., source mismatch, selection functions of the surveys, measurement calibration, or contamination.

\begin{figure}
  \begin{center}
    \includegraphics[width=0.99\columnwidth]{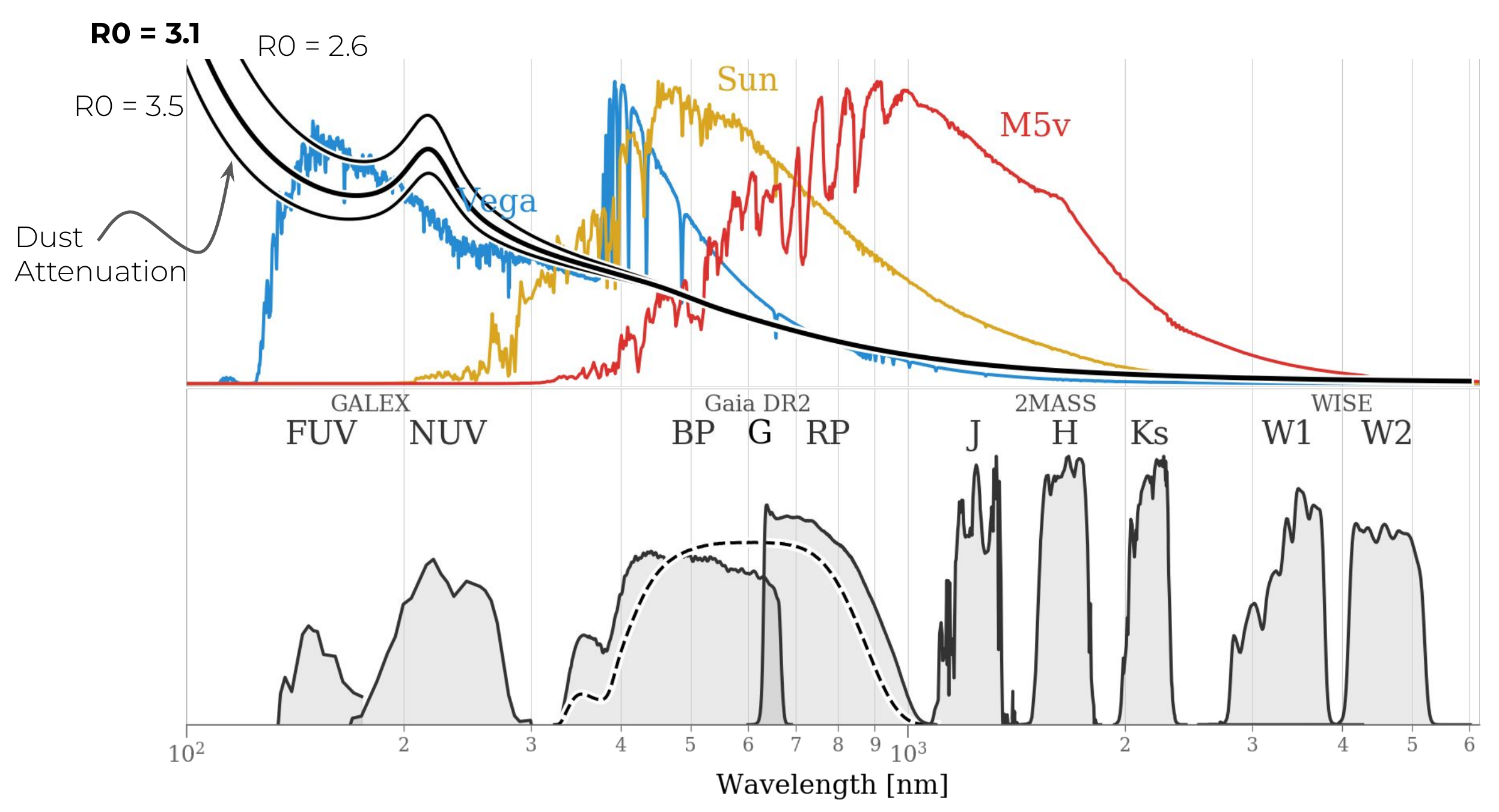}
  \end{center}
  \caption{Photometric filters covering the various compiled datasets
  (details in \autoref{sec:datasets}) compared with spectra of typical stars:
  Vega (A0V), a G2V star (Sun-like star), and an M5III star.
  The bottom panel shows the GDR2 transmissions of the three Gaia passbands and the selected all-sky survey ones (2MASS, and WISE). The top panel shows selected spectral templates from \citet{Pickles1998} and we overlaid the \citet{Fitzpatrick1999} {dust extinction curve} and its variations with $R_0$ for reference.
  We include GALEX for reference, but we did not include the survey in this work (see Sect. \ref{sec:datasets}).
  }
  \label{fig:filters}
\end{figure}

We constructed a crossmatch catalog by considering the ``obvious'' all-sky non-Gaia data provided by DPAC\footnote{DPAC: (Gaia) Data Processing and Analysis Consortium.} that span a different wavelength range, i.e., 2MASS \citep{2MASSsurvey, 2MASSCat} and AllWISE \citep{AllWiseCat}. For the latter, Wise provides us with four photometric bands spanning wavelengths from 1 to 30 $\mu m$.
Because the interstellar dust emits at those wavelengths contaminating W3 and W4 photometry, we only included W1 and W2 in our analysis.
We initially considered including GALEX \citep{GalexGR5cat} in our analysis. As shown in Fig.\,\ref{fig:filters}, GALEX would offer an essential advantage to analyze blue and hot stars as well as helping us in inferring dust properties and chemical patterns \citep[e.g.][]{Kaviraj2007}. However, DPAC did not provide us with a crossmatch between Gaia and GALEX. Furthermore, the GALEX catalog possesses complex photometric systematics and a significantly broad point spread function. As a result, constructing a reliable crossmatch is a significant endeavor on its own -- for instance, forced photometry approaches such as in \citet{Lang2014, NeoWiseR3}.
Our selected photometric bands cover the wavelength range from $\sim 300$nm to $\sim 5\,{\mu}m$.  Special attention went into the passbands and the various photometric/magnitude systems present across the different surveys (see \autoref{sec:models} for details).

Typical reported photometric uncertainties for Gaia, 2MASS and AllWise are reported in Table\,\ref{tab:inputcatuncert}. At first glance, it may look plausible to have uncertainties of $\sim 0.1$\,mag, i.e., a few percent in flux measurements. However, a significant fraction of sources has lower reported uncertainties than $0.05$ mag, which is not reasonable if one accounts for external calibration uncertainties (e.g., knowledge of the passbands). When combining photometric surveys, reported uncertainties may not be accounting for all the limits of the calibration. For our analysis, we altered uncertainties to be at least  $0.05$\,mag ($\max(\sigma, 0.05)$; see Table\,\ref{tab:inputcatuncert}).

The GDR2 photometry and parallaxes are also not free from systematic errors \citep{Lindegren2018, Lindegren2018a, MaizApellaniz2018, Zinn2019}. Various studies (including the DPAC release papers) detail prescriptions to apply to the Gaia data to account in part for these systematics. In the present work, we applied the calibration rules summarized in Table\,\ref{tab:calibtable}. The recalibrated values are also reported in our catalog (see Appendix\,\ref{sec:catalog}).

Let us also emphasize that we consider here crossmatch catalogs with very different astrometric accuracies. We are merging information from surveys with varying coverages of wavelengths and magnitude limits. Therefore where we obtain incorrect matches between surveys, sources will inevitably have corrupted SEDs despite the careful compilation from \citet{Marrese2019}. This issue will occur most often in dense stellar regions (e.g., clusters) and regions with high dust content (bulge, star-forming region). In Sect.\,\ref{sec:models}, we describe how we rendered our model more flexible to account for (additional) unknown systematics.

\begin{table}
    \centering
    \caption{Typical uncertainties in magnitudes of the (raw) input photometric catalogs: minimum and a few quantiles.}
    \begin{tabular}{c|c|c|c|c|c}
    \hline
    Catalog & Band & min        & 25\%        & 50\%      & 75\% \\  
    \hline\hline
            & BP   &  $10^{-8}$ &  $0.0046$   & $0.0113$  & $0.0309$ \\ 
     Gaia   & G    &  $10^{-5}$ &  $0.000613$ & $0.00105$ & $0.0019$ \\ 
            & RP   &  $10^{-8}$ &  $0.00237$  & $0.0051$  & $0.0106$ \\ 
    \hline
            & J    &  $0.013$   &  $0.032$    & $0.056$   & $0.101$\\ 
     2MASS  & H    &  $0.010$   &  $0.036$    & $0.070$   & $0.13$\\  
            & Ks   &  $0.011$   &  $0.041$    & $0.0885$  & $0.16$\\  
    \hline
    AllWise & W1   &  $10^{-3}$ & $0.026$     & $0.034$   & $0.044$\\ 
            & W2   &  $10^{-3}$ & $0.035$     & $0.062$   & $0.117$ \\ 
    \hline
    \end{tabular}
    \label{tab:inputcatuncert}
    \end{table}

\begin{table*}
\caption{Prescriptions applied to the input data and references.}
\label{tab:calibtable}
\begin{center}
\begin{tabular}{c|c|l|l}
\hline
Parameter & Intervals & Calibrations & References \\
\hline\hline
                     & $\gmag < 14$    &  $\varpi + 0.05$ mas  & \citet{Lindegren2018a, Zinn2019}  \\
 $\varpi$  & $14 \leq \gmag < 16$ &  ${\varpi} + (0.1676 - 0.0084\cdot \gmag)$ mas       &  \citet{Lindegren2018a}, (interpolation)  \\
                     & $ 16 \leq \gmag$  &   $\varpi + 0.029$ mas  & \citet{Lindegren2018}  \\
\hline
                    &    $G < 11$    &   $1.2 \cdot \sigma_\varpi$  & \\
 $\sigma_{\varpi}$  &    $11 \leq G < 15$  &   $(0.22\cdot \gmag - 1.22) \cdot \sigma_\varpi$      &  \citet{Lindegren2018a}  \\
                   &        $ 15 \leq G$  &   $\left(\exp(15 - \gmag) + 1.08\right) \cdot \sigma_\varpi$  &    \\
\hline
     &    $ \gmag < 6$         & $\gmag + 0.0271 \cdot (6 - \gmag)$  & \\
 $G$ &    $6 \leq \gmag < 16$  & $\gmag - 0.0032 \cdot (\gmag-6)$ &  \citet{MaizApellaniz2018}  \\
     &    $16 \leq \gmag$      &   $\gmag - 0.032 $  &  \\
\hline
 \bpmag  &    $\gmag < 10.5$    &   bright $\bpmag$ transmission curve   & \citet{MaizApellaniz2018} \\
 \bpmag  &    $ 10.5 \leq \gmag$  &   faint $\bpmag$ transmission curve   & \citet{MaizApellaniz2018} \\
\hline
$\sigma_{\rm mag}$  & Gaia, 2MASS, WISE  & $\max\{\sigma_{\rm mag}\}, 0.05 {\rm mag}\} $  &    \\
\hline
\end{tabular}
\end{center}
Details are provided in Sect. \ref{sec:datasets}
\end{table*}

In \autoref{sec:catalog}, we describe the catalog content.
The final input catalog contains a total of \ntotcat\ sources having Gaia parallaxes and all eight photometric bands from Gaia, 2MASS, and AllWISE.

\begin{figure}
    \begin{center}
      \includegraphics[width=0.8\columnwidth, clip, trim=0 0 0 0em]{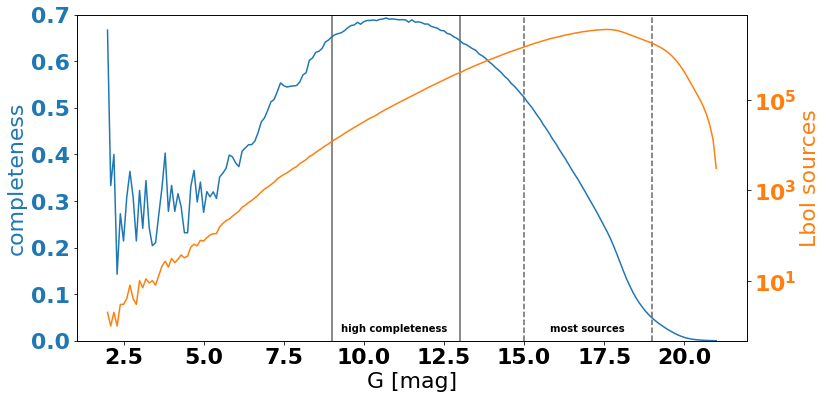}\\
      \includegraphics[width=0.8\columnwidth, clip, trim=0 0 0 0em]{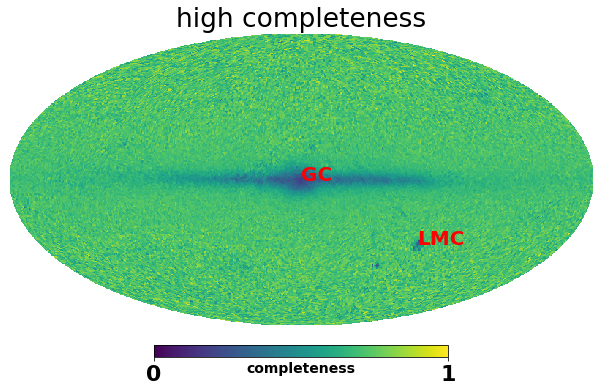}\\
      \includegraphics[width=0.8\columnwidth, clip, trim=0 0 0 0em]{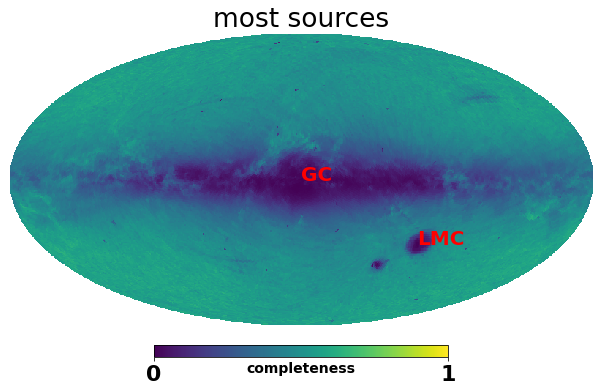}
    \end{center}
    \caption{Completeness of our astrophysical parameter catalog with respect to all Gaia DR2 sources.
    Top panel shows the completeness and the Lbol sourcecount as a function of magnitude G in blue and orange, respectively.
    In the middle (lower) panel the completeness over the sky (in Mollweide projection and Galactic coordinates) are shown for the high completeness (most sources) magnitude range indicated in the top panel with the solid (dashed) vertical lines, i.e. for sources with 9$<$G$<$13 (15$<$G$<$19).
    }
    \label{fig:catcompleteness}
  \end{figure}

Fig.\,\ref{fig:catcompleteness} shows the completeness of our catalog w.r.t. \gdr{2}.
We can see that regions of high star density are the most incomplete due to the substantially lower spatial resolution than Gaia of the 2MASS and AllWise surveys.
While including non-Gaia data into the analysis will certainly help improve the results, non-Gaia data are not available for all the stars.
One may hope in the future for forced photometry catalogs of 2MASS and AllWISE using Gaia (possibly GALEX).
Nevertheless, dedicated Gaia-only studies such as \citet{GDR2Andrae2018} have merit in their own right. They analyze self-consistent data, report about their limitations, and they set a baseline for further analysis.

\section{Dust attenuated stellar models and assumptions}\label{sec:models}

Various authors have described approaches to extract stellar properties using models trained on empirical and synthetic SED data (e.g., support vector machine, forward-modeling, neural-networks). We do not need to re-investigate these methods. Instead, we first briefly detail our adopted models and our fitting procedure.

We define a model that predicts the {\it dust attenuated magnitudes} in the same photometric bands as our data $\left\{M_k\right\}_{k\in[1..K]}$ given a set of APs $\pars$ that fully defines a star for our model. This first includes its atmosphere $\left(\log_{10}(T_{\rm eff}), \log_{10}(L/L_\odot),  {\rm [Fe/H]}\right)$, temperature, luminosity, and metallicity, respectively; and age to uniquely map the evolution stage, $\log_{10}({\rm age/yr})$.
{One could rightfully argue that age, (initial) mass, and metallicity should be sufficient to set the stellar SED uniquely (e.g. stellar tracks, isochrones). An (age, mass, [Fe/H]) model grid requires the sampling density variations to span multiple orders of magnitude to represent stellar evolution fairly. The choice of (age, mass, [Fe/H]) generates a suboptimal grid structure to support our accurate interpolation scheme (e.g., Andrae et al. 2022, {\it in prep}). Instead, we adopt the atmosphere parameters that directly translate the SEDs and spread more evenly the grid points, but to the expense of an additional intrinsic dimension to uniquely tie the grid points to the stellar evolution.}
However, we eventually marginalize over the metallicity dimension. We discuss this point in Sect. \ref{sec:metallicity}.
Additionally, we account for the effect of the interstellar medium (ISM) on the observed spectrum of the star through two extinction parameters, $\left(\Azero, \Rzero\right)$;
Afterward, we can compute the final photometry with the eight passbands mentioned above. Our parameters additionally include the distance modulus $\mu$ to the star and our noise model jitter, $\log_{10}\jitter$.

We detail the various ingredients of this model below and the prior assumptions we use in our inference.

\paragraph*{Evolution models} --
For this work, we generated an extensive collection of models by combining the PARSEC isochrones (PARSEC 1.2s + Colibri (PR16), \citealt{Chen2014, Marigo2013, Rosenfield2016}) and the ATLAS9 atmosphere models \citep[$3500 < \teff < 50\,000$]{Castelli2004}.  We had to recompute the spectra to apply the dust prescription detailed below adequately.
{We did not include non-LTE models in this analysis for two reasons: (i) adopting the same atmospheric models allowed us to directly validate our models against the PARSEC predictions and (ii) overall variations of $\sim 0.01$\,mag between LTE and non-LTE models on integrated photometry (e.g., \citealt{Mitchell2017} across various wavelength ranges.}
We do not include pre-main sequence, post thermally pulsating asymptotic giant branch (TP-AGB), and white-dwarf stars in our model grid.  The two former evolution phases are intrinsically rare, and the UV and NIR predictions are very model-dependent for these three evolution stages, respectively.
The final set of evolution grid points covers log(A/yr) between $6.6$ and $10.2$\,dex ($0.2$ dex step) from and [M/H] between $-2$ to $0.2$\,dex ($0.2$\,dex steps). The isochrones define the sampling along the mass dimension to represent all evolution stages regardless of their durations.

\paragraph*{Dust prescription} -- We must account for the fact that photons travel from the star to the observer through the interstellar medium (ISM). Dust along the line of sight will absorb or scatter some photons. As a result, the star's light will be dimmer (similarly to a distance effect) and redder than its intrinsic color. We adopted a model of dust {extinction} following \citet{Fitzpatrick1999}. This prescription depends on \Azero and \Rzero, the dust {extinction} at $\lambda=550$\,nm, and the average {extinction} per color excess unit ($\Rzero=3.1$ is the average value in their model for the Milky-Way, also referred to as the average grain size parameter). {The literature also refers to these parameters at $A_V, R_V$. We adopted the notations from the Gaia Consortium that aims at avoiding confusions with the effects on integrated photometry.}
We apply the dust extinction to the nominal spectra for every point from the evolution model while varying the \Azero and \Rzero values. Our grid spans \Azero from $0$ to $20$\,mag ($0.2$\,mag step) and \Rzero from $2.5$ to $3.7$ ($0.3$ step).
Only then, we compute the photometry of the individual spectra through the relevant passbands. As discussed in detail in \citet{Gordon2016} for an example, one must apply the dust before the filters to avoid non-linear effects.

\paragraph*{Model interpolation scheme} -- For our application of fitting the integrated magnitudes with model predictions computed from synthetic spectra, we must keep in mind that most of the computation cost comes from the generation of the models.
As we also approach this problem with MCMC sampling, we need to generate the models on the fly for every source at a given $\pars$. The computation cost using a traditional interpolation scheme (e.g., multi-linear) becomes rapidly expensive for a single source.
Instead, our solution resides in ``emulating'' our stellar models: we replace our grids of SED models with an accurate representation that is fast to evaluate. To do so, we adopted a Multivariate Adaptive Regression Splines \citep[MARS;][]{Friedman1991} \footnote{
\href{https://github.com/scikit-learn-contrib/py-earth}{Python} or \href{https://cran.r-project.org/web/packages/earth/index.html}{R} packages use the name ``earth''.}.
This approach is similar to a simplified neural net with ReLU activation functions.
This approach is very similar to a simplified neural net with ReLU activation functions. However, there are differences from a neural network approach.
First, the model construction uses simple principles: making a local regression model in the stellar model space; each bin in the parameter space has a set of equations to preserve continuity between the bins.
Second, the model is rigid, i.e., limiting the use of arbitrary functions; thus, overfit is less significant in this approach, and we obtain very mathematically smooth predictions.
Finally, the model's interpretability allows us to validate the model locally and globally rapidly; the final model is a series of extremely fast analytic expressions to evaluate.
MARS models are essentially analytic spline equations, which are much faster to evaluate than real-time grid interpolation. The speed performance gain is significant compared with multi-dimensional interpolations or Support Vector Machine, for instance.

\paragraph*{Noise model and bad SEDs} -- Corrupted SEDs and systematics may affect the photometric measurements (Sect. \ref{sec:datasets}). We include in our likelihood a photometric \textit{jitter} noise model that allows us to inflate the uncertainties on a source by source basis. More specifically, we define per star $\jitter$ as a common quantity across all the bands that we add in quadrature to all of our photometric uncertainties (on top of the prescriptions in Table\,\ref{tab:calibtable}).
For the band $k$, the probability density of the observed data given the parameters, i.e., the likelihood is
\begin{align}
    & \proba{m_k \given m, \sigma_k, \jitter} \\
    & = \frac{1}{\sqrt{2\pi(\sigma_k^2 + \jitter^2)}}\exp\left(-\frac{1}{2}\frac{(m_k - m)^2}{(\sigma_k^2 + \jitter^2)}\right),\nonumber
\end{align}
where $(m_k, \sigma_k)$ are the magnitude measurements, and $m$ is the true (unknown) magnitude value.
The normalization of the Gaussian is the key that ensures $\jitter$ cannot arbitrarily increase: an optimization procedure will compromise between the exponential and the term in front of it.

The jitter term captures issues in our data assuming these are random effects on the magnitudes, in contrast with systematic ones coming from incorrect zeropoints. Photometric or parallax offsets are fully degenerate with distance, and there is not much we can do in our star-by-star inference, unfortunately.

{We did not consider a jitter per survey or even per band. Preliminary tests showed that the individual survey jitter parameters attempt to equalize the signal to noise of each survey, especially affecting the Gaia bands. While this could help capturing more systematics, this increases the fitting complexity with highly non-linear terms, and reduces the gain from high precision photometry.}

\paragraph*{Likelihood per source} -- For every source, we consider the set of the $K=8$ photometric measurements $\left\{m_k\right\}_{k\in[1..8]}$ and the parallax $\varpi$ (all with Gaussian uncertainties $\sigma_{x}$). Given the model above the log-likelihood of a star given a set of APs $\pars$, a distance modulus $\mu$ and jitter $\jitter$:
\begin{align}
    -2 \,\cdot\, & \ln\proba{\left\{m_k\right\}_{k\in[1..8]}, \varpi \given \pars, \jitter}\\
    = & \ln({2\pi\sigma_\varpi^2}) + \frac{1}{\sigma_\varpi^2}\left(\frac{\varpi}{1\mas} - \frac{1\kpc}{r}\right)^2 +
    \nonumber\\
     & \sum_{k=1}^{K}\left[\ln\left({2\pi(\sigma_k^2+\jitter^2)}\right) + \frac{\left(m_k - M_k(\pars) - \mu \right)^2}{\sigma_k^2 + \jitter^2}\right].
    \nonumber
\end{align}

\begin{table*}
\caption{Model priors.\label{tab:priors}}
\begin{center}
\begin{tabular}{c|l|l}
\hline
Parameter & Prior Distribution & Notes \\
\hline\hline
 \Rzero    &  $\Rzero \sim \mathcal{N}(3.1, 0.2)$ & Gaussian on standard value; \citet{Fitzpatrick1999}\\
 \Azero\ / \mags    &  $\Azero \sim \exp(-|\Azero|))\times \mathcal{U}[-0.1, 20]$ & exponential over $[-0.1, 20]$\,mag interval \\
 \hline
 \dmod\ / \mags    & $\dmod \sim \mathcal{U}[-5, 19]$ & uniform distance modulus from 1\,pc to 60\,kpc\\
 \hline
 *\feh\ / dex    & $\feh \sim  \mathcal{U}[-2, 0.2]\times \mathcal{N}(\mu_{A}, \sigma_{A})$ & Age dependent Gaussian over model range\\
 & $\quad\mu_{A} = 0; \sigma_{A} = 0.25$ & $\quad$ if $\log_{10}(A/yr) < 9$ \\
 & \quad $\mu_{A} = -3 \log_{10} (A) ; \sigma_{A} = 0.5 \log_{10} (A) + 0.25$ & $\quad$ if $\log_{10}(A/yr) \geq 9$ \\
 \begin{tabular}{@{}c@{}}\teff\ / K \\ $\log(L / L_\odot)$\end{tabular} &$\rm T_{eff}, \log(L) \sim \mathcal{U}[HRD]$ & uniform in the region covered by isochrones\\
 \hline
 $\log_{10}(A/yr)$ & $\mathcal{U}[6, 13]$ & uniform log-age over the isochrone range \\
 **$\rm M / M_\odot$ & ${\rm M} \sim {\rm IMF}(0.01, 120\,M_\odot)$  & \citet{Kroupa2001} initial mass function. \\
 \hline
 $\log_{10}(\jitter / mag)$ & $\jitter \sim \mathcal{N}^+(0, 0.01)\times \mathcal{U}[0, 0.3]$  & Half-Gaussian truncated at $0.3$\,mag. \\
 \hline
\end{tabular}
\label{calibtable}\\
\end{center}
Details are provided in Sect. \ref{sec:models}\\
* [Fe/H] is prior dominated and marginalized in our posterior (see \autoref{sec:metallicity})
\\
** the mass is not a sampling parameter but predicted by our model.
\end{table*}

\paragraph*{Priors}
The more stringent prior assumptions in your analysis, the more tensions between your models and the data will exist. In this present work, we avoid detailed assumptions such as the Milky-Way 3D stellar population structures. Using less informed priors should not alter the results when the data are "good" -- the (multi-dimensional) likelihood sufficiently narrow and when the photometry agrees with the parallax. When the likelihood distribution is broader, weak priors do not impose our knowledge on the data allowing for potential discoveries.

Table \ref{tab:priors} summarizes our prior assumptions. We detail these choices below.

$\bullet$ The dust extinction prior allows for substantial values $\Azero$ (up to $20$\,mag). We also allow slightly negative values to avoid edge effects with the MCMC sampling. $\Rzero$ is very likely to vary with environment \citep[e.g.][]{Cardelli1989, Bianchi1996, Valencic2004, Gordon2009} but there is no model of these variations yet. We assume variations to be around the average value of $\Rzero=3.1$ from \citet{Fitzpatrick1999} with a standard deviation of $0.2$ (which includes the $2.7$ mentioned in the cited paper).

$\bullet$ Our distance prior is uniform in distance modulus between $[-5, 19]$\,mag which corresponds to a range of $1\pc$ and $70\kpc$. The upper distance limit allows us to include the Magellanic Clouds ($\sim40-50\kpc$) and some space to account for outliers and spurious parallaxes.

$\bullet$ We include isochrones in our model.
Therefore we impose an Hertzsprung Russell Diagram (HRD) in the allowed range of temperatures and luminosities. We also enable the metallicity to vary over the model range from $-2$ to $0.2$\,dex. However, we found that we do not constrain the metallicity with our analysis (see Sect. \ref{sec:metallicity}). {This prior is a constant Gaussian on solar value with 0.2\,dex dispersion up to 1\,Gyr, which then shifts to a mean of -1 dex and 0.5\,dex dispersion at 10\,Gyr (see right panel fig.\,\ref{fig:metallicity-prior-dominate})}. We consider a uniform distribution of logarithmic ages from $6$ to $13$\,dex, {bounded by} our isochrone selection.
However, to avoid over-predicting very massive stars with a lot of extinction, we adopt the initial mass function (IMF) from \citet{Kroupa2001} and use the stellar mass our model predicts.

$\bullet$ Finally, our noise model parameter is assumed as small as possible through a Gaussian prior centered on zero. We also impose $\jitter$ to be smaller than $0.3$\,mag, which we set to avoid significant outliers.

It is essential to highlight that the above priors do not have any particular Galactic position dependence (incl. distance). As we allow more flexibility in general, we expect more significant uncertainties than studies with a Milky-Way model prior \citep[e.g., StarHorse,][]{Anders2017, BailerJones2018}.

\section{Validation}\label{sec:validation}

{We sampled the posterior distribution over $\{\rm log_{10} T_{eff}, log_{10}(L), A_0, R_0, \mu, {\rm [Fe/H]}, \log_{10}(age)\}$ for $123\,097\,070$ stars observed by Gaia, 2MASS, and AllWise}. We now assess the quality and limitation of our results including predicted \logg\ values {using statistical diagnostics and literature reference catalogs}.

{
The reader will find detailed analysis below, but briefly we highlight our main findings. We find a very good agreement with the distances from \citet{BailerJones2021}, with 2 to 8\% absolute fractional differences within 5\,kpc. One can trace these differences to our distance prior.
Our results may overall overestimate the extinction by $\sim0.1$\,mag by comparing against literature catalogs and samples of Red Clump stars. This may result from a lack of strong extinction prior in our analysis. We also qualitatively find that our $\Rzero$ maps agree with previous work (limited overlaps) and the variations correlate.
Our temperatures appear biased high by $\sim 300$\,K w.r.t. literature values. They are in better agreement with the StarHorse catalog than the values we compiled from various surveys.
Our $\logg$ values agree with asteroseismology with a median absolute difference below 0.3\,\dex.
We find that our ages, distances and extinction estimates agree well for wide binaries.
}

\subsection{CMD/HRD}

\begin{figure*}
    \begin{center}
        \includegraphics[width=1.8\columnwidth, clip]{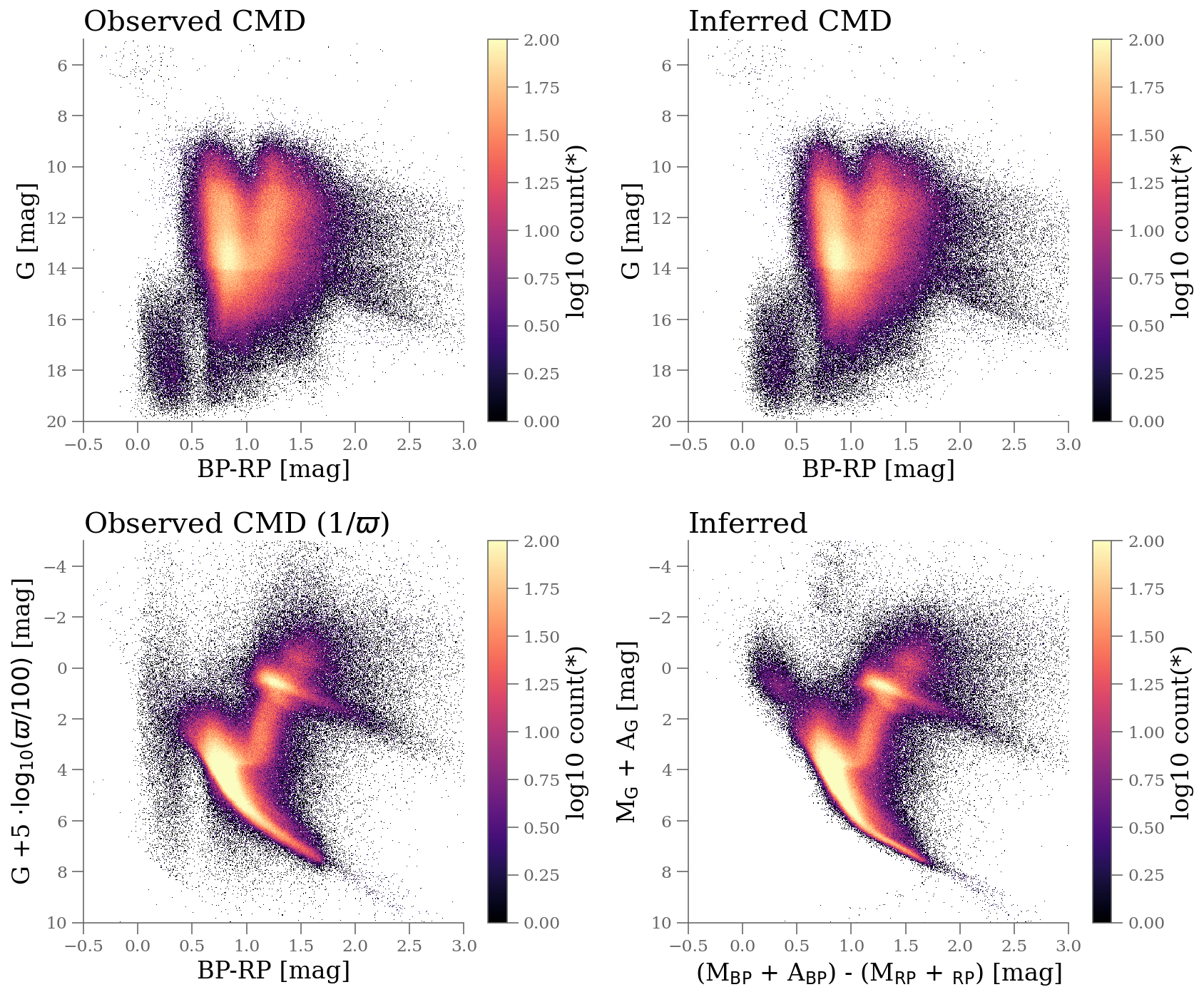}
      \end{center}
      \caption{Overview our analysis procedure on \textit{validation sample} containing 853,610 stars. Top panels present observed color-magnitude diagrams where left and right panels are the input data and their median predictions, respectively. The lower panels show the inverse parallax distance corrected CMD and the distance corrected obtained from the AP estimates. The quantity on the y-axis of these two panels would be identical in absence of parallax noise. The entire catalog is shown in appendix Fig.~\ref{fig:cmdcheck} {and the corresponding residuals between the top panels Fig.\ref{fig:cmdcheck_residuals}}.
      }
      \label{fig:vstcmdcheck}
\end{figure*}

Figure \ref{fig:vstcmdcheck} presents an overview of the observables and their predictions from our inference. We also report the inferred photometry of each source as part of our inference outputs, which allows us to compare the observed and model color-magnitude diagrams (CMDs; top panels). These are nearly identical. The distance corrected ones (lower panels) show a few more discrepancies: the lower main-sequence seems thinner, our inference leads to many blue sources being hot stars. Nevertheless, overall, we obtain a robust agreement. We also emphasize that we see around the red clump the RGB and AGB bumps on the distance corrected CMDs (lower right panel). We show the same figure for the entire catalog in appendix Fig.~\ref{fig:cmdcheck}.

\begin{figure}
    \begin{center}
        \includegraphics[width=\columnwidth, clip, trim=19cm 0 0 0]{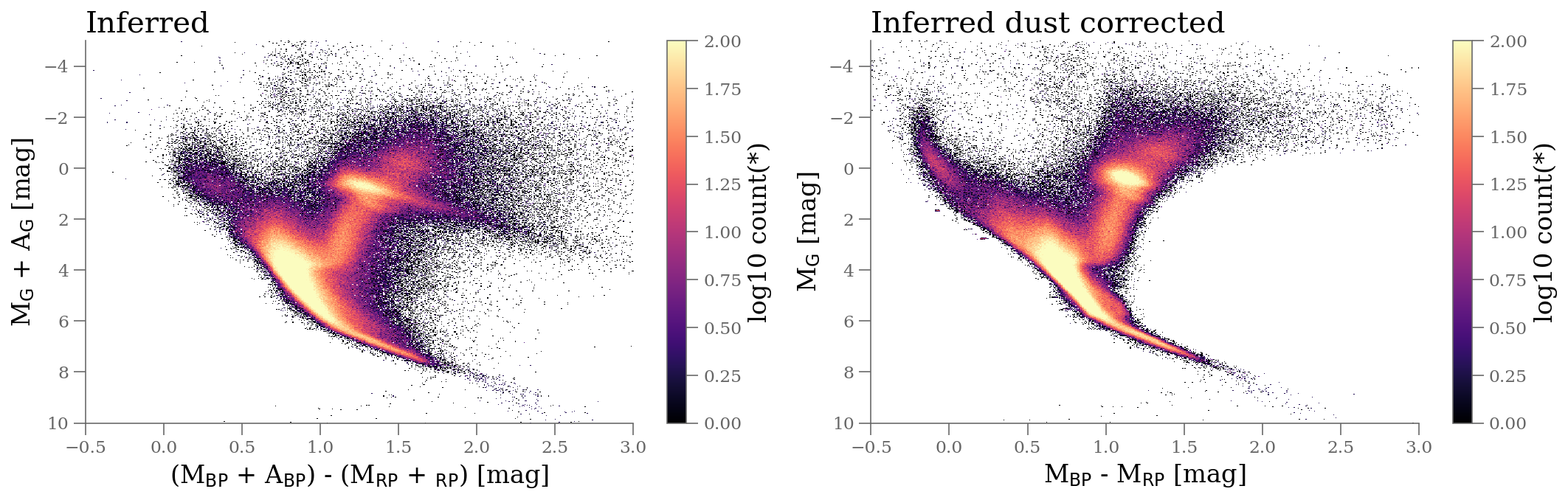}
      \end{center}
      \caption{
        Inferred color absolute magnitude diagram (CAMD) after accounting for the dust extinction.
        (The CAMD before accounting for the dust corresponds to the lower right panel of Fig.\,\ref{fig:vstcmdcheck}.). The entire catalog is shown in appendix Fig.~\ref{fig:cmddustcheck}.
      }
      \label{fig:vstcmddustcheck}
\end{figure}

\begin{figure*}
    \begin{center}
        \includegraphics[width=2\columnwidth, clip]{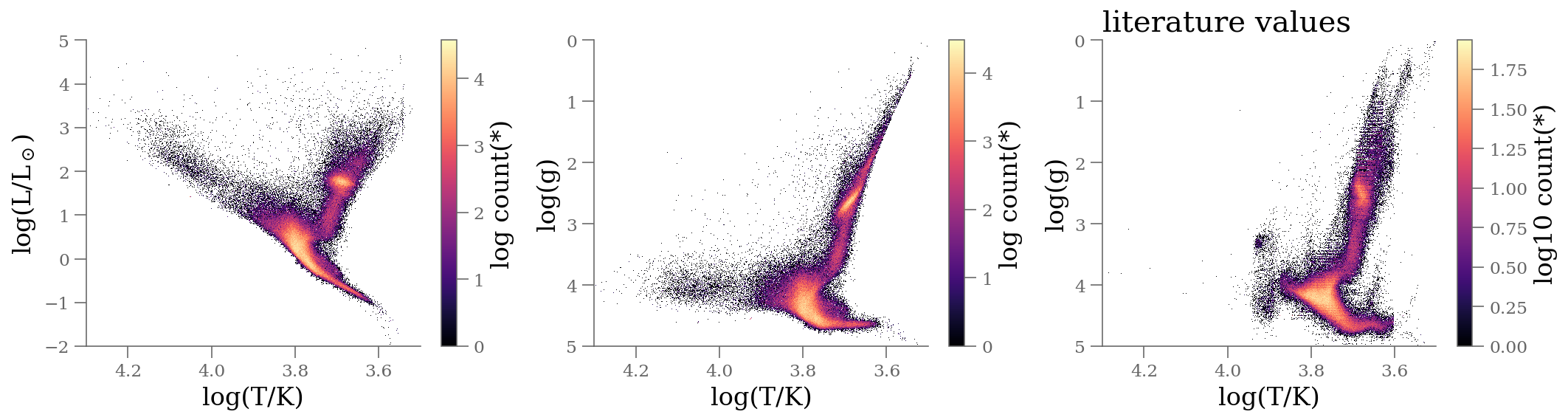}
      \end{center}
      \caption{
        HRD and Kiel diagrams of the validation sample (853,610 stars). On the right-hand side panel, we show the literature values. The features on the graphs are smooth, and the red clump is narrow. One could also see the Blue helium-burning giants.
      }
      \label{fig:vst_hrd}
\end{figure*}

Figure\,\ref{fig:vstcmddustcheck} shows our results after removing the inferred dust contribution. It compares with the lower right panel of Fig.\,\ref{fig:vstcmdcheck}. Based on the significant change of the red clump extend after removing the dust component, our inference reduces the extinction effects. No obvious spurious feature appears after correcting for extinction effects in the CAMD. The hot stars create the usual hook, although we note that it is not confirming these are real hot stars. (If their distances are also in agreement with their parallax values, this could be a confirmation; see below). The apparent gap at the top of the main sequence results from the sample selection. All the stars on the right-hand side of the giant branch move back to the giant and main sequence. The AGB sequence (RGB tip) is well populated. We also note the over-density on the left side of the RGB above the red-clump: these are He-burning stars. We show the same figure for the entire catalog in appendix Fig.~\ref{fig:cmddustcheck}.

We do not find significant issues in these results so far. We now look at the HDR and Kiel diagrams of these sources as shown in Fig.\,\ref{fig:vst_hrd}. For reference, the right-hand side panel shows the Kiel diagram made from the literature values. On all these diagrams, all features are smooth: thin main-sequence, giant branch, clear red-clump. We note that the literature values have issues with the hot stars and a double giant sequence. Also, we note that the literature shows a wiggle in the MS, which cannot be physical. Overall, the agreement is excellent.

\subsection{Distances}

\begin{figure*}
    \begin{center}
        \includegraphics[width=2\columnwidth, clip]{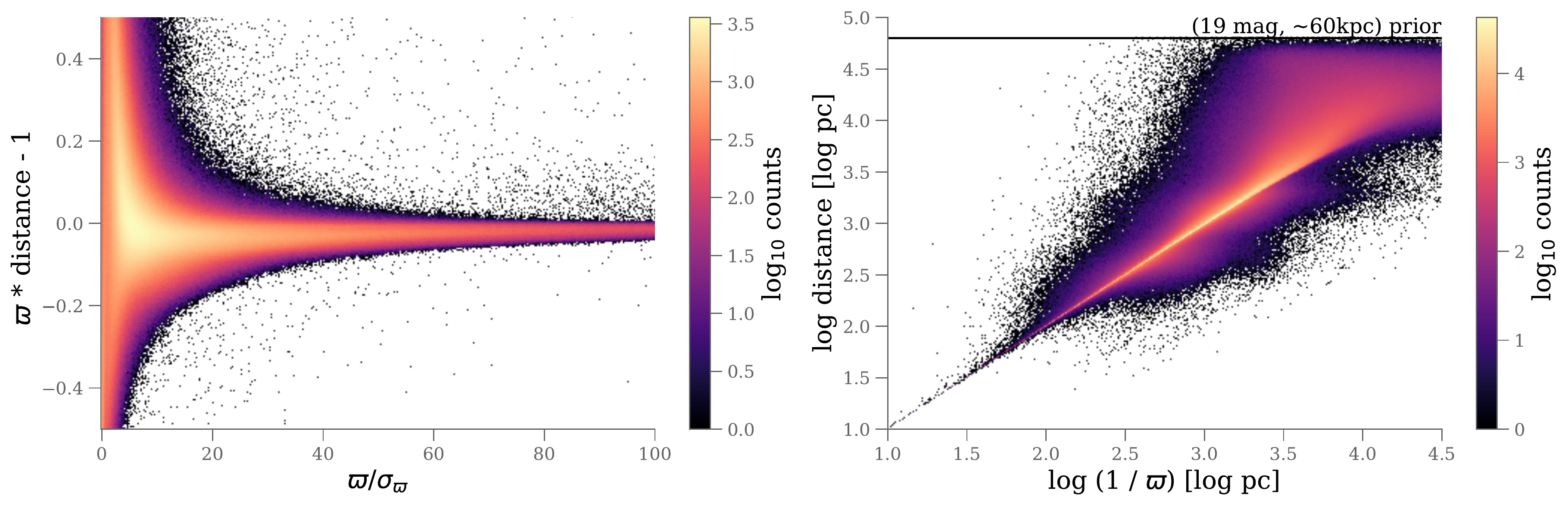}
      \end{center}
      \caption{
        Distance estimates w.r.t. the input parallaxes of the entire catalog. Left shows the parallax distance product as a function of the parallax signal-to-noise ratio. Right shows the distance vs. inverse parallax distribution for positive parallax objects. The distance saturation on this panel is due to our distance modulus prior (19 mag).
      }
      \label{fig:vst_distance_vs_parallax}
\end{figure*}

Figure \ref{fig:vst_distance_vs_parallax} compares our distance with the Gaia parallaxes. In the left panel, we plot the product of distance and parallax as a function of parallax SNR. We corrected for the unity offset to center the distribution on 0. The more precise the parallax, the tighter our distance estimates as expected. A bias is visible when parallax uncertainties are large. This trend comes from the significant number of negative parallaxes (below 0). For these sources, we must infer a positive distance and within our Galaxy. Many of these sources also have values inconsistent with null parallaxes, making the distance difficult to reconcile. The right panel shows our estimates when selecting positive parallaxes only (regardless of SNR). The agreement is excellent. Note that we doubt objects with inverse parallax further than $30$\,kpc ($sim$ upper limit of the x-axis) to be actual single stars from our Galaxy.

\begin{figure*}
  \begin{center}
   \includegraphics[width=1.0\textwidth]{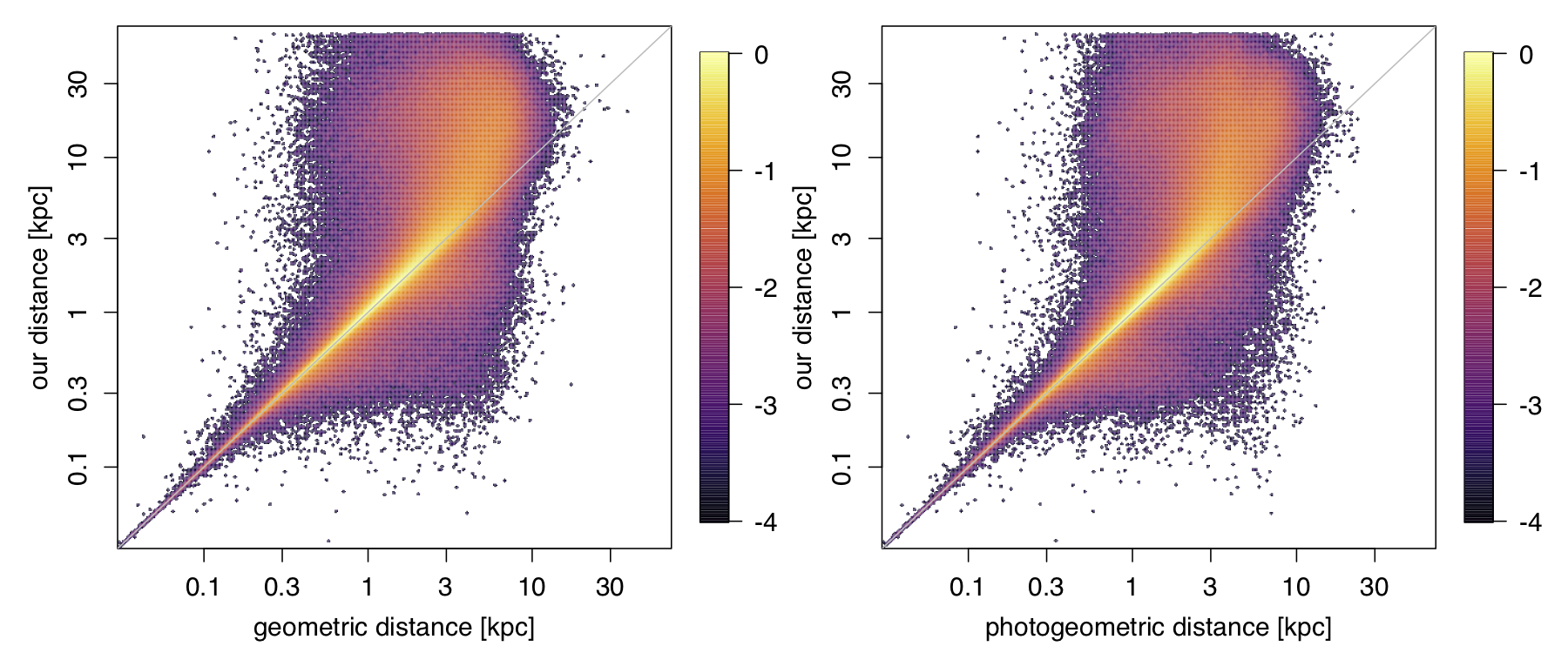}

  \end{center}
  \caption{Comparison of the median distance in our catalog (vertical axis) with the median geometric (left) and median photogeometric (right) distances from the Gaia EDR3-based catalog of \cite{BailerJones2021}. The color scale indicates the density of sources in each panel on a log10 scale relative to the maximum. Note the logarithmic distance axes. The diagonal line is the identity line.
  }
  \label{fig:paperV_dist50_vs_rmed}
\end{figure*}

\cite{BailerJones2021} (hereafter BJ2021) inferred distances for all stars in Gaia EDR3 that have parallaxes (including negative ones). These geometric distances used the parallax together with a direction-dependent distance prior, which they defined using a Galaxy model. For those stars with BP-RP colors, they
additionally inferred their photogeometric distances, which use this color along with the G magnitude and an HRD-prior to keep distances consistent with stellar models.
Figure~\ref{fig:paperV_dist50_vs_rmed} compares our distances with both of these estimates, using a million stars randomly selected in common to both catalogs. Both their study and ours provide posterior medians, so a direct comparison is possible. (Comparing a median to a mode or mean would introduce a bias). We observe excellent agreement up to 1\,kpc, which is not surprising because the high precision parallaxes dominate the distance inference within this range.
At further distances, more significant differences occur due to the increasing importance of photometric information in our estimates and the difference in the methods' priors.
For the whole sample, the median fractional difference with respect to the geometric distances, defined as $(r_{\rm ours} - r_{\rm BJ2021})/r_{\rm BJ2021}$, is $+0.07$. This is a measure of the bias. A measure of the scatter is the {\em absolute} median fractional difference,  $|r_{\rm ours} - r_{\rm BJ2021}|/r_{\rm BJ2021}$, which is $0.15$ for the geometric distances.
Comparing our distances to the photogeometric distances, these two measures are
$+0.10$ (bias) and $0.15$ (scatter).
The scattering amplitude is reasonable given the median fractional parallax uncertainty (fpu), $|\sigparallax/\parallax|$, is $0.20$ for this sample (recall that all parallaxes in the comparison are positive).
Recall also that we use Gaia DR2 parallaxes whereas \cite{BailerJones2021} use Gaia EDR3 parallaxes, which has 20\% more precise parallaxes on average.
There are several possible causes of the bias, not least the different distance priors and additional assumptions about the relationship between color and absolute magnitude (i.e., the underlying stellar models). As we use a weaker distance prior than was adopted in \citet{BailerJones2021}, our estimates can extend to much larger distances. When limiting the comparison to sources with $|r_{\rm ours} <5$\,kpc (median fpu is $0.13$), we find median and median absolute fractional differences of $0.02$ and $0.08$ relative to the geometric distances (respectively), and $0.03$ and $0.08$ relative to the photogeometric distances.
So clearly, the bias is dominated by distant sources.

\subsection{Dust extinction \Azero, \Rzero}\label{sec:dust}

\paragraph{Red-Clump Stars} --
We first compare our distances to red-clump star samples. The absolute luminosities of stars in the red-clump are reasonably independent of their chemical composition, mass, or age. For instance, \citet{Stanek1998} measured the absolute magnitudes for the red-clump at solar metallicity at $-0.22$ mag in the I-band with a variance of about $0.15$ mag.
Therefore red-clump stars are good proxies to isolate dust reddening and extinction effects, as argued e.g.~in \citet{GDR2Andrae2018}.

\begin{figure}
    \begin{center}
        \includegraphics[width=\columnwidth, clip]{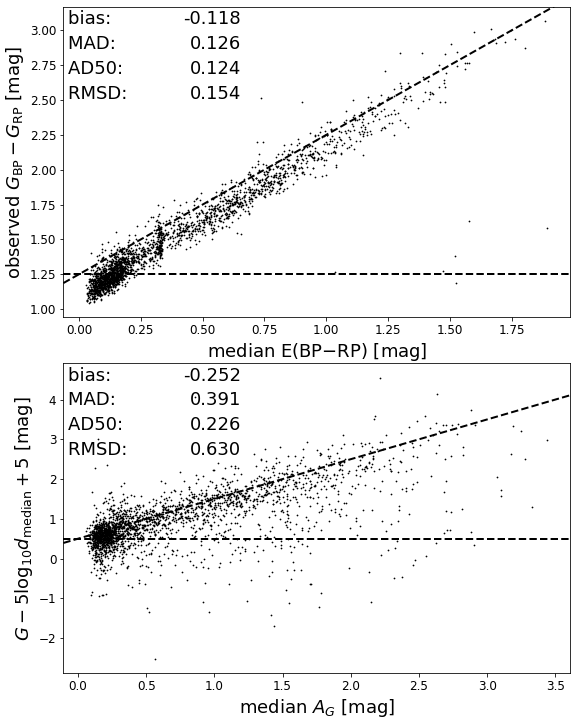}
      \end{center}
      \caption{Comparison of color vs. our median reddening estimates (top panel) and reddened absolute magnitude vs.~our median $A_G$ estimates (bottom panel) for red-clump stars from \citet{Bovy2014}. Horizontal dashed lines indicate the intrinsic color and magnitude, whereas diagonal dashed lines indicate the aspired identity relation.
      }
      \label{fig:red-clump-reddenings-median}
\end{figure}

In Fig.~\ref{fig:red-clump-reddenings-median}, we compare our reddening and extinction estimates to the colors and magnitudes of red-clump stars from \citet{Bovy2014}.
Our reddening estimates (top panel) are in excellent agreement, whereas the lower panel of this figure suggests some additional scatter coming from either our $A_G$ or our distance estimates.

\begin{figure*}
  \begin{center}
      \includegraphics[width=2\columnwidth, clip]{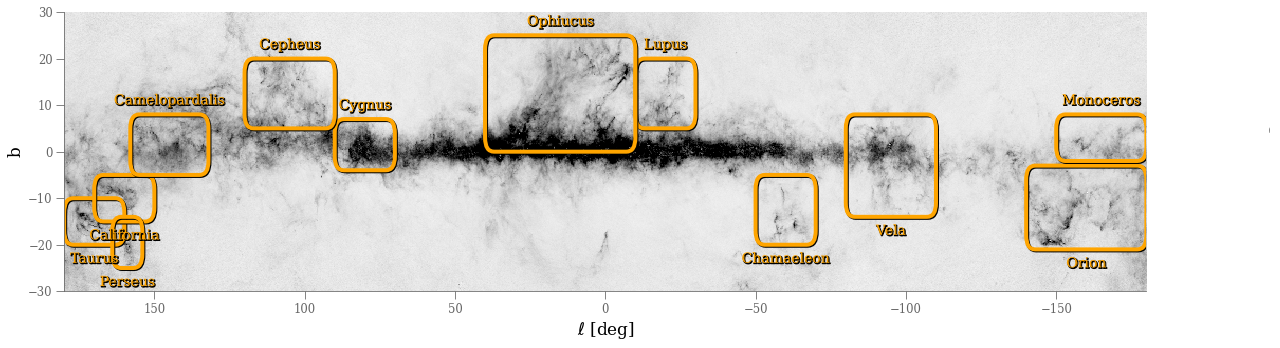}\\
      \includegraphics[width=2\columnwidth, clip]{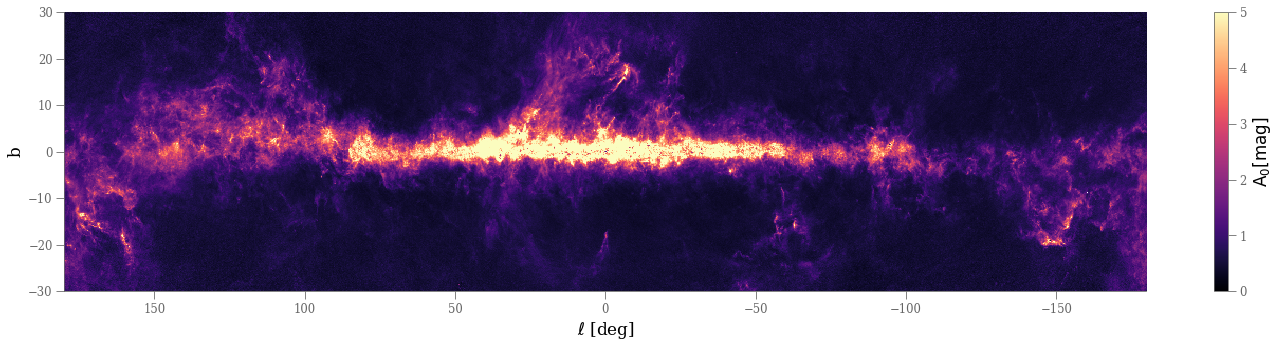}\\
      \includegraphics[width=2\columnwidth, clip]{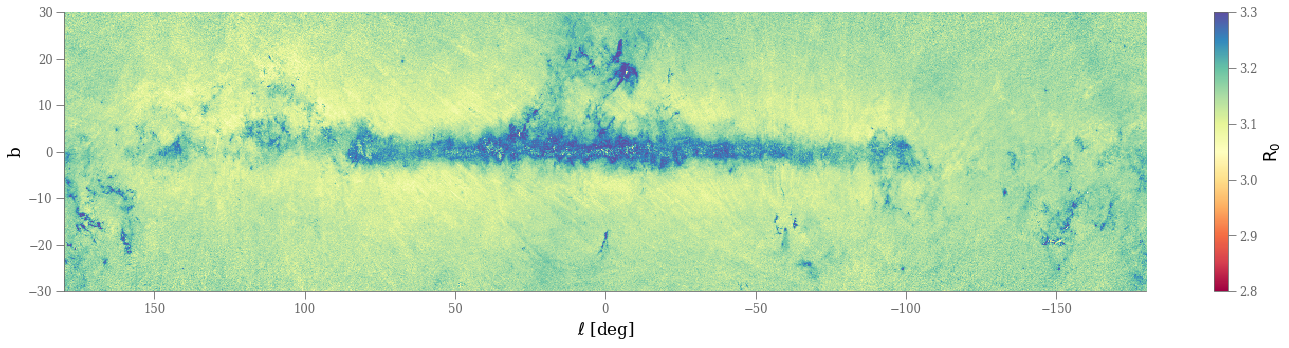}
    \end{center}
    \caption{Sky distribution in Galactic coordinates (averaged over all distances) of the dust extinction parameters \Azero (middle) and \Rzero (bottom).
    We indicate some molecular regions of our Galaxy by the rectangles on the top panel (overlay of the Gray scaled \Azero map).
    The maps are centered on the Galactic Center, with longitudes increasing towards the left.
    We only plot the 60 degrees centered on the Milky-Way disk in these panels.
    Highlighted regions are zoomed in in appendix Figs.~\ref{fig:mw_a0_r0_maps_zooms} \& \ref{fig:mw_a0_r0_maps_zooms_2}.
    }
    \label{fig:mw_a0_r0_maps}
\end{figure*}

The statistical validity of our $\Azero$ and $\Rzero$ estimates is further attested to by Fig. \ref{fig:mw_a0_r0_maps}. Recall that we do not use any sky position during our inference: each star estimate remains independent of any other. This plot shows sharp features distinct between the two panels and significantly different from just plotting the Gaia color.
The sky maps of our extinction estimates highlight the complexity of the Milky Way disk's ISM. We recover a wealth of features across a wide range of scales, from thin filaments to large cloud complexes. The Perseus, Taurus, and Orion complexes dominate the anti-central region (far left and right sides of the map, respectively), and the Ophiuchus molecular cloud complex (above the Galactic Center) shows exquisite substructures.
Using our extinctions and distances, \citet[in press]{Dharmawardena2021}  inferred the structure of the Orion, Taurus, Perseus, and Cygnus X star-forming regions. They locate Cygnus X at 1300--1500~pc, in line with VLBI measurements. They concluded our catalog would support studies of the changes in grain size or composition of dust as processed in molecular clouds.

\begin{figure}
    \begin{center}
        \includegraphics[width=\columnwidth, clip]{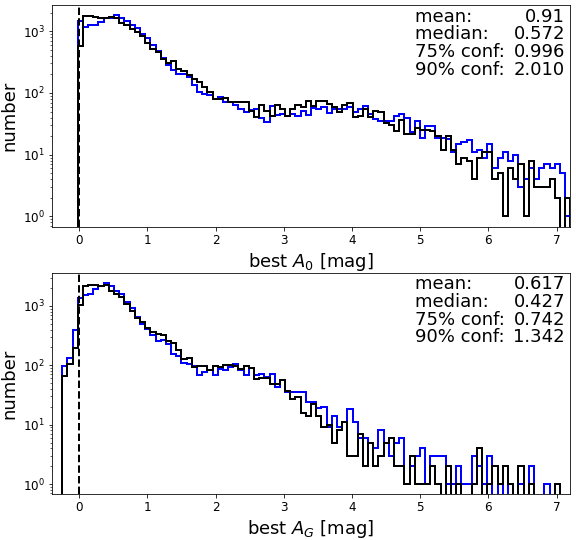}
      \end{center}
      \caption{Distribution of \Azero\ (top panel) and $A_G$ (bottom panel) estimates in the Local Bubble ($\varpi>20$\,mas and $\varpi/\sigma_\varpi > 5$) for our median values (black histogram) and maximum posterior values (blue histogram). Numbers quote \Azero\ statistics from our median values.
      }
      \label{fig:AG-Local-Bubble}
\end{figure}

\paragraph{Local Bubble} -- We also investigate the distribution of our extinction estimates for stars in the Local Bubble, i.e., stars having $\varpi>20$mas ($\sim 50$\,pc) and good parallax measurements ($\varpi/\sigma_\varpi > 5$). In this very spatially nearby sample, we do not expect any significant extinction \citep[e.g.,][]{Vergely2010}. Fig.~\ref{fig:AG-Local-Bubble} shows that results in the Local Bubble have virtually no differences between median and maximum posterior (``best'') estimates. However, our extinction values appear overall too high for this sample: half of the stars have $A_0>0.572$\,mag, and 25\% of stars even $A_0>0.996$\,mag. This trend may result from the median values of truncated distributions may not be the most adapted statistics \citep[e.g.,][]{GDR2Andrae2018}, or spurious astrometry \citep{Rybizki2021}. It may well be the result of the weak priors we adopted.

\begin{figure}
    \begin{center}
        \includegraphics[width=\columnwidth, clip]{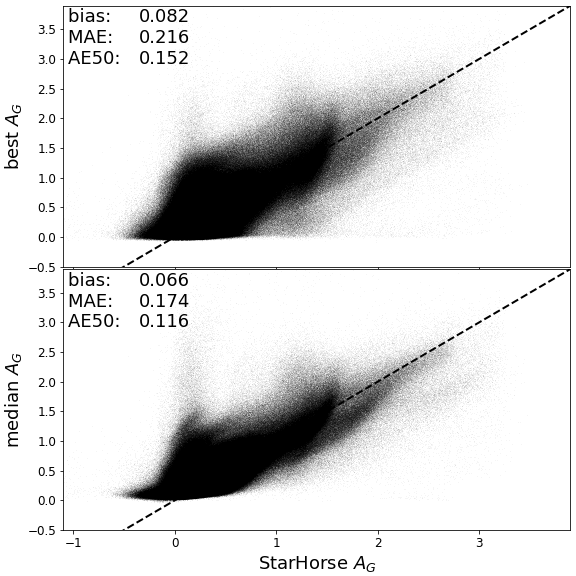}
      \end{center}
      \caption{Differences of $A_G$ extinction estimates between our best result (top panel) and our median result (bottom panels) to values from \citet{Anders2017} (they provide median statistics). Numbers quote various statistics to summarise the differences.
      }
      \label{fig:Lbol-AG-vs-StarHorse}
\end{figure}

\paragraph{StarHorse Catalog} -- Finally, we compare our extinction estimates $A_G$ to those of \citet{Anders2017}, a catalog that similarly inferred properties of stars using Gaia data. These values in both methods are not a constant scaling from the models' $\Azero$ parameters but account for the shape of the stellar spectrum. Overall,  Fig.~\ref{fig:Lbol-AG-vs-StarHorse} shows a good agreement between the two sets despite the estimates differ in their constructions. As \citet{Anders2017} provided median estimate values, it is not surprising that our median $A_G$ values agree slightly better than our maximum posterior estimates with those of StarHorse.

\begin{figure*}
    \begin{center}
        \includegraphics[width=\textwidth, clip, trim= 1cm 1cm 1cm 1cm]{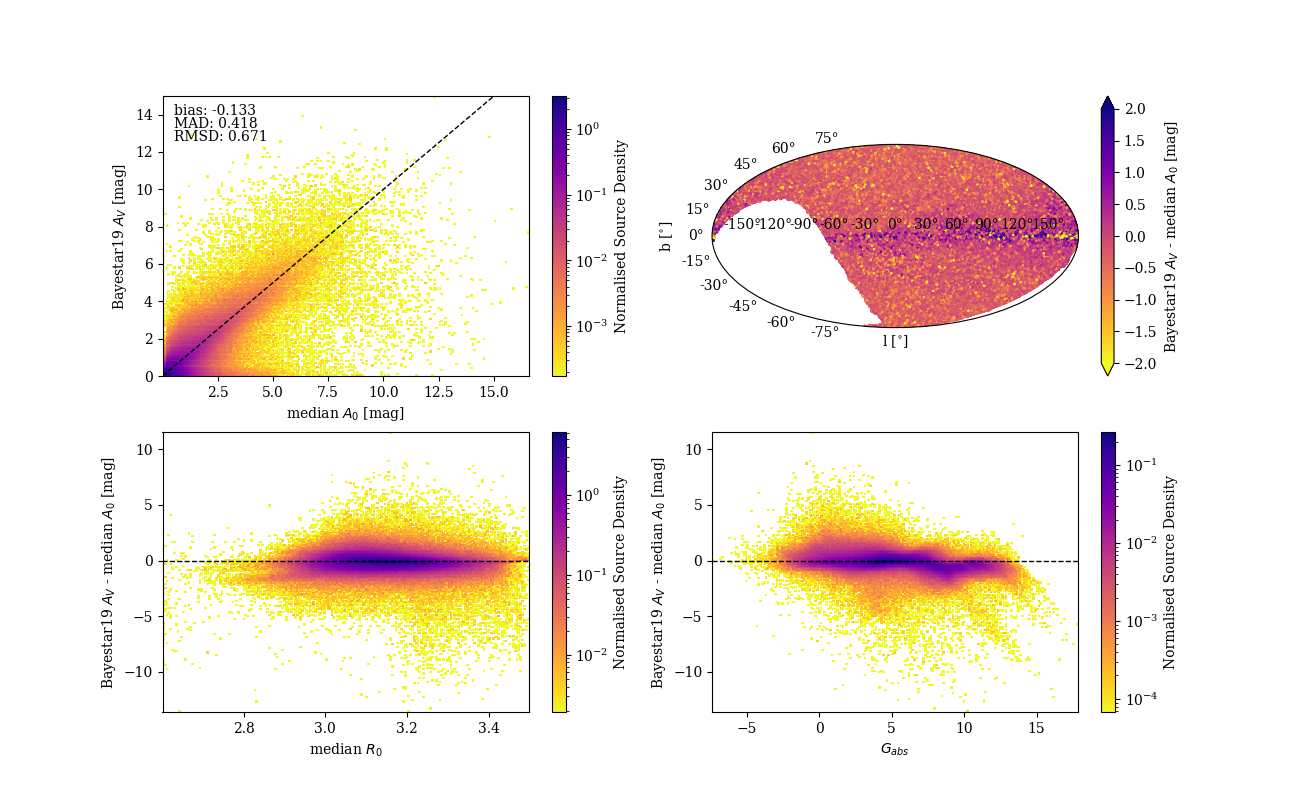}
      \end{center}
      \caption{Comparison of our extinctions (median \Azero) to the Bayestar19 catalog A$_{V}$ for a random sample of 925\,527 sources in common. The dashed black lines in each plot represent 1:1 line. Top left: Our median \Azero\ compared to Bayestar19 A$_{V}$; top right: residuals of Bayestar19 A$_{V}$ - Our median \Azero\ as a distribution on sky in Galactic coordinates; bottom left: same residuals compared to our median \Rzero\ estimates; bottom right: same residuals compared to our median estimate of intrinsic absolute magnitude G$_{abs}$}
      \label{fig:BayeStarComp_RandSources_1mil_combined}
\end{figure*}

\paragraph{BayeStar Catalog} -- In figure \ref{fig:BayeStarComp_RandSources_1mil_combined}, we compare our extinctions (median \Azero) to the Bayestar19 catalog extinctions (A$_{V}$, \citealt{Green2019}) for a random sample of $\sim 1$ million sources in common to both catalogs.
For $62\%$ of the random sources, we find that our median \Azero\ values are larger than the estimates from Bayestars19, with a mean difference of $0.1$\,mag. Given that Bayestars19 predicts extinction with finite (grid) resolution while we do not impose such a constraint, the systematic difference may reflect unresolved structures for the Bayestars's approach.

Figure \ref{fig:BayeStarComp_RandSources_1mil_combined} also compares the extinctions residuals between the two catalogs w.r.t. our \Rzero\ and absolute magnitudes. There is no significant correlation of the residuals with \Rzero. In contrast, the comparison with absolute magnitudes suggests we slightly overestimate extinction at low luminosity, i.e., the lower part of the main sequence. This may result from our weak prior in comparison to Bayestar's.

{We compared our \Rzero values to those presented in \citet[their Fig. 1]{Schlafly2017}. The distributions agree qualitatively. A further quantitative comparison remains difficult, as their definition differs from the standard and the limited accessibility to their values. We will continue to investigate this issue in the future.}

\subsection{Temperature, gravity, mass, radius}\label{sec:stellar_aps}

To further validate our stellar parameters, we created a sample of reference by compiling literature values from SDSS/APOGEE \citep{ApogeeDR14}, GALAH \cite{Buder2018}, Gaia-ESO \citep[iDR5]{Gilmore2012}, LAMOST \citep{2011RAA....11..924W, 2014IAUS..306..340W} and RAVE \citep{2017AJ....153...75K, 2013AJ....146..134K}.
Details on the filtering on each of the catalogs are given in \ref{appendix:VST-definition}.

\begin{figure*}
    \begin{center}
        \includegraphics[width=2\columnwidth, clip]{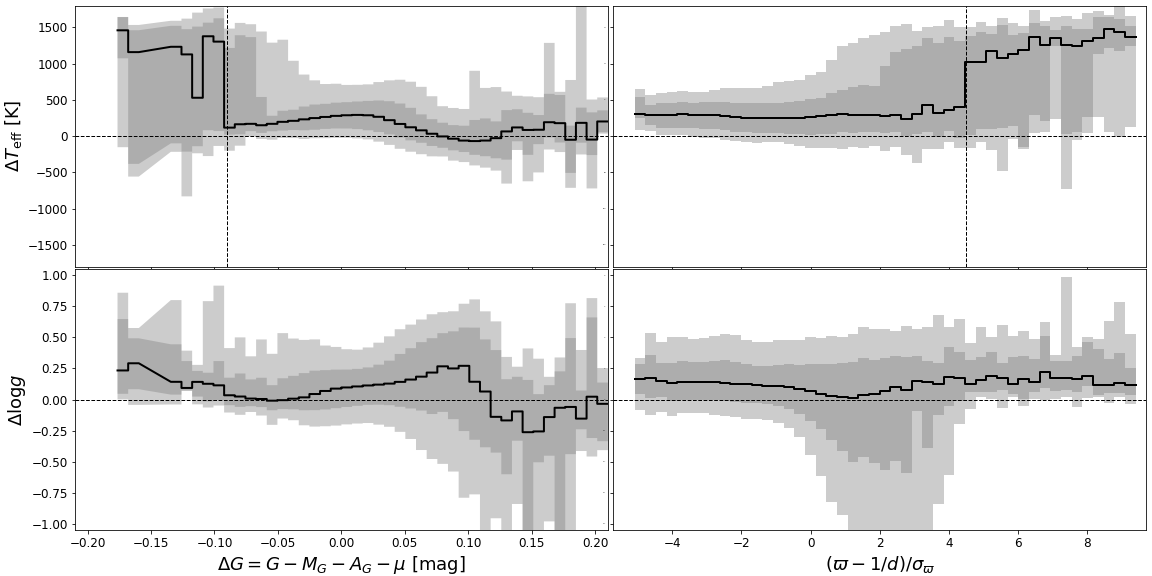}
      \end{center}
      \caption{Residuals of \teff\ (top panels) and \logg\ (bottom panels) w.r.t.~literature reference sample (Sect. \ref{sec:stellar_aps}; Appendix~\ref{appendix:VST-definition}) as functions of fit residuals for apparent $G$ (left panels) and parallax $\varpi$ (right panels). The solid lines indicate the median values in each bin and the shaded regions are the central 68\% and 90\% intervals, respectively.
      }
      \label{fig:Lbol-AP-quality-vs-fit-residuals}
\end{figure*}

We compare the residuals of \teff\ and \logg\ to the reference sample in Fig.~\ref{fig:Lbol-AP-quality-vs-fit-residuals}. This figure shows residuals on both axis: the differences to literature values for \teff\ and \logg\ on the y-axis and the apparent $G$ magnitude (left panels) and the parallax (right panels) on the x-axis, respectively.
While \logg is relatively well behaved throughout the sample, the \teff\ differences rise sharply for fit residuals below $\Delta G<-0.09$ and above $\frac{\varpi-1/d}{\sigma_\varpi}>4.5$ (indicated by the vertical dotted lines on the plots). However, this affects less than $0.2$\% of our results. Unfortunately, there is also a systematic overestimation of \teff by $\sim$300K, which is also visible on the top-right panel in Fig.~\ref{fig:Lbol-AP-quality-vs-fit-residuals}.

\begin{figure*}
    \begin{center}
        \includegraphics[width=2\columnwidth, clip]{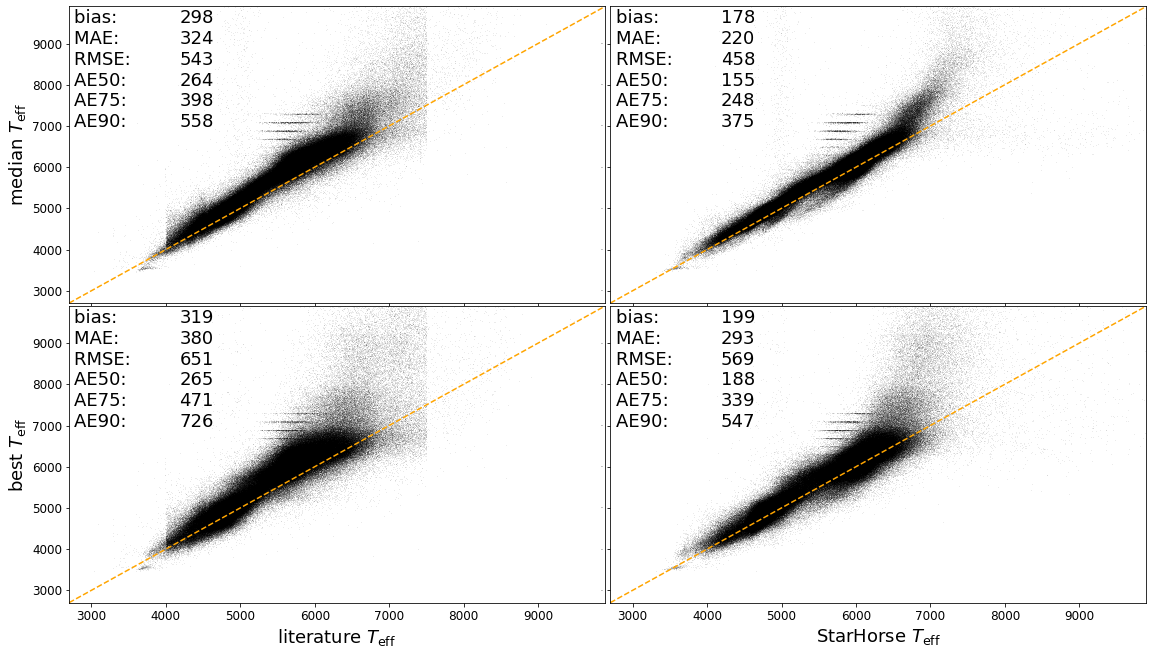}
      \end{center}
      \caption{Differences of median \teff\ (top panels) and maximum posterior \teff (best; bottom panels) to literature values (Appendix~\ref{appendix:VST-definition}, left panels) and StarHorse values form \citet{Anders2017} (right panel). In each panel we quote several statistics summarising the differences.
      }
      \label{fig:Lbol-Teff-vs-literature}
\end{figure*}

Figure~\ref{fig:Lbol-Teff-vs-literature} compares further our \teff\ estimates (median and maximum posterior) with the literature sample and the StarHorse catalog \citep{Anders2017}. This figure shows that our median values compare substantially better to literature temperatures than our ``best'' values. Unfortunately, it confirms a systematic overestimation of \teff\ by $\sim$300K (left panels). However, our \teff\ estimates are substentially more in agreement with the StarHorse ones, with a departure for hot stars ($\teff> 7000$\,K). It is unclear if the bias is due to our weak priors or a stellar model mismatch with the data. The larger discrepancies with literature may be the result of combining various heterogeneous catalogs.

\begin{figure}
    \begin{center}
        \includegraphics[width=\columnwidth, clip]{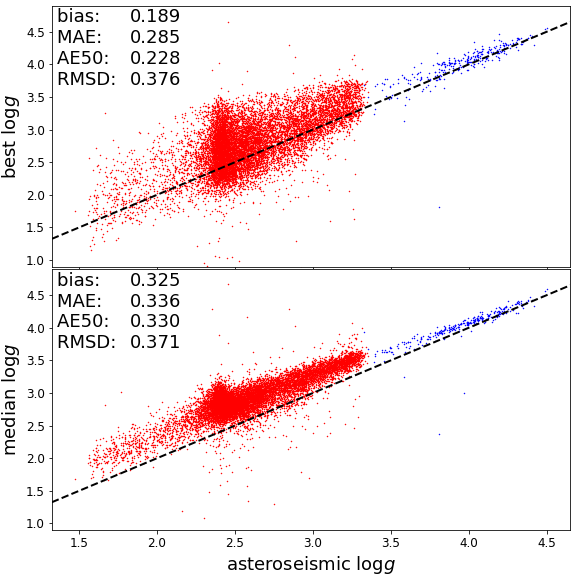}
      \end{center}
      \caption{Differences of our maximum posterior (top panel) and median (bottom panels) \logg\ estimates to asteroseismic values from \citet{Serenelli2017} (main sequence stars; blue points) and \citet{Yu2018} (giant stars; red points).
      }
      \label{fig:EDR3-Lbol-logg-asteroseismology}
\end{figure}

We also compare our \logg\ estimates to asteroseismic values from \citet{Serenelli2017} and \citet{Yu2018}. From Fig.~\ref{fig:EDR3-Lbol-logg-asteroseismology}, our best \logg values compare well to asteroseismic values having median absolute differences below $0.3$\dex. However, our approach appears to overestimate the median \logg values, especially for giant stars (red points).

\subsection{Wide binaries}\label{sec:widebinaries}

\begin{figure*}
    \begin{center}
        \includegraphics[width=2\columnwidth, clip]{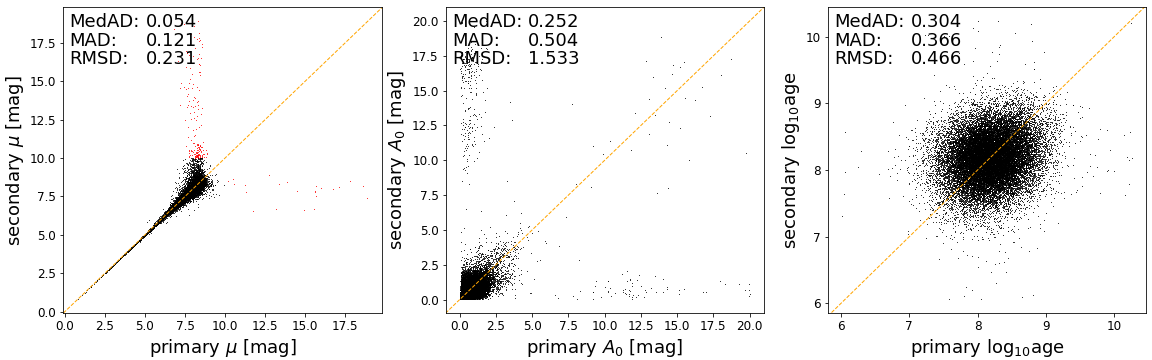}
      \end{center}
      \caption{Wide binaries from \citet{WideBinaries} and a comparison of distance moduli (left panel), $A_0$ (middle panel) and log-ages (right panel) for primary and secondary. In the left panel, red points mark pairs where one component is further away than 1kpc. Such cases are excluded from the number statistics and from all other panels.
      }
      \label{fig:wide-binaries-validation}
\end{figure*}

We also considered a sample of wide binaries from \citet{WideBinaries} where Gaia observes both components individually.
These systems are commonly co-eval, making them one of the best testbeds to assess the quality of distances, extinctions along the lines of sight, and ages of field stars.
Figure~\ref{fig:wide-binaries-validation} compares the distance moduli, extinction \Azero, and $\log_{10}(age)$ of both components. The distance moduli and also the log-ages agree very well between the components. However, the extinction estimates (middle panel of Fig.~\ref{fig:wide-binaries-validation}) are spuriously large for one component but not both, in some cases. The discrepancy could result from our weak priors but could also likely indicate where the  crossmatch between surveys went wrong, resulting in producing incorrect input SEDs for our analysis. Nevertheless, the bulk of the wide binary pairs in the sample agree within $0.2$\,mags.

\subsection{Uncertainties}\label{sec:uncertainties}

Assessing the quality of reported uncertainties is always a challenge. We must assume that the reference datasets are unbiased and that their uncertainties are correctly calibrated. In this section, we first use the above wide binary sample, which allows us to check the internal consistency of our estimates. Then we use the literature sample define above for further assessment.

\begin{table}[h]
\caption{Number of instances where the central 68\% and 50\% confidence intervals do \textit{not} overlap between components of 27\,125 wide binaries taken from \citet{WideBinaries}}
\label{tab:uncertainties-wide-binaries}
\begin{center}
    \begin{tabular}{l|r|r}
         parameter & 68\% intervals  & 50\% intervals  \\
         \hline
$\mu$ &  3950 ($\sim$14.6\%)  &  8246  ($\sim$30.4\%) \\
$A_0$ &  5531 ($\sim$20.4\%) &  9585  ($\sim$35.3\%) \\
$\log_{10}$age &  4 ($\sim$0.01\%) &  11 ($\sim$0.04\%) \\
    \end{tabular}
\end{center}
We give statistics for the distance modulus $\mu$, extinction \Azero, and log-age. The 50\% confidence interval is ``smaller'' than the 68\% interval such that we expect more cases of non-overlap. We restricted the sample to pairs where both components are within 1\,kpc and $\Azero < 5$\,mag to exclude the spuriously high extinctions.
\end{table}

\paragraph{Using Wide Binaries} -- We use the previous sample of wide binaries of \citet{WideBinaries} (Sect.\,\ref{sec:widebinaries}) and investigate to what extent the confidence intervals overlap for distance moduli, extinctions, and ages that should be similar in both components. Table~\ref{tab:uncertainties-wide-binaries} gives the number of cases where the 68\% and 50\% confidence intervals do \textit{not} overlap. Unfortunately, even if we assume that literature values and our estimates are unbiased or have identical systematic errors, we cannot predict how often this should happen.%
\footnote{
  Assuming Gaussian uncertainties, one could estimate the probability that two estimated values are outside each other's confidence intervals for each source individually but not for the whole sample. Yet, our uncertainties are certainly not Gaussian.
}
Instead, we have to rely on these numbers being ``plausible''. We report these statistics in Table~\ref{tab:uncertainties-wide-binaries}. The ``mismatch'' seems to happen far too often for the distance modulus ($\mu$) and extinction ($\Azero$), and virtually never for our age estimates (which is not a sampled parameter). These numbers could be reflecting underestimated uncertainties or the MCMC converging to local minima that are inconsistent with a binary system. It could happen that some pairs from \citet{WideBinaries} not being genuine binaries at all. However, Fig.~\ref{fig:wide-binaries-validation} suggests that the solutions of both components are in excellent agreement with each other consistent with most pairs from \citet{WideBinaries} very likely being real binaries. Table~\ref{tab:uncertainties-wide-binaries} thus suggests that we systematically underestimate our uncertainties for distance and $A_0$. For log-age, the results from Table~\ref{tab:uncertainties-wide-binaries} suggest that the uncertainties are too large, leading to too few cases where the intervals do not overlap.

\begin{figure}
    \begin{center}
        \includegraphics[width=\columnwidth, clip]{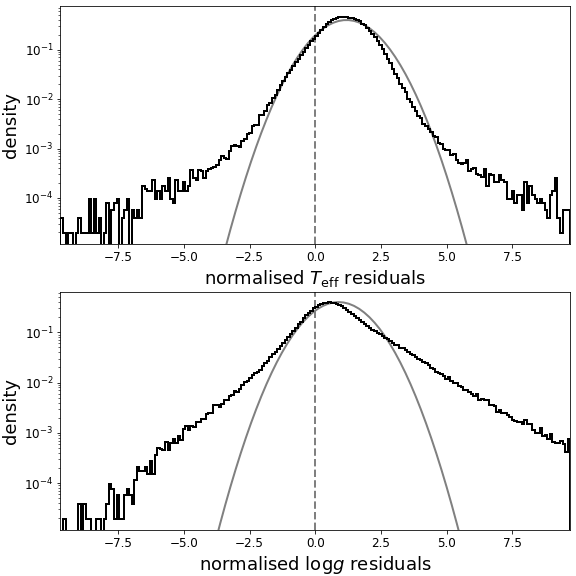}
      \end{center}
      \caption{Distribution of normalised residuals to literature values for Teff (top panel) and logg (bottom panel). Grey curves show a Gaussian having the same mean and unit standard deviation for comparison.
      }
      \label{fig:Lbol-normalised-residuals-Teff-logg}
\end{figure}

\paragraph{Literature Reference catalog} -- we inspect the distribution of differences between our \teff\ and \logg\ estimates and literature values (sample described in Sect.\ref{sec:stellar_aps}).  Fig.~\ref{fig:Lbol-normalised-residuals-Teff-logg} shows this distribution normalized by the symmetrized 68\% confidence interval and the quoted literature error (note the log-scale on the y-axis). These normalized residuals do not peak at zero, thus indicating biases. We already noted the $\sim 300$\,K in Sect. \ref{sec:stellar_aps}. Otherwise, they are reasonably close to a Gaussian with unit standard deviation (for \logg, the tails are a bit heavier than a unit Gaussian). One can attribute further departure from the unit Gaussian to the non-Gaussianity of the posterior distribution (multivariate or marginalized). This comparison with spectroscopic values implies that at least for \teff\ and \logg\ our 68\% confidence intervals are reasonable estimates of the \textit{random} errors in our results.\footnote{Systematic errors are generally not accounted for in confidence intervals since those characterize only \textit{random} errors.}

\subsection{Metallicity}\label{sec:metallicity}

\begin{figure*}
    \begin{center}
        \includegraphics[width=2\columnwidth, clip, trim=0 0 0 1cm]{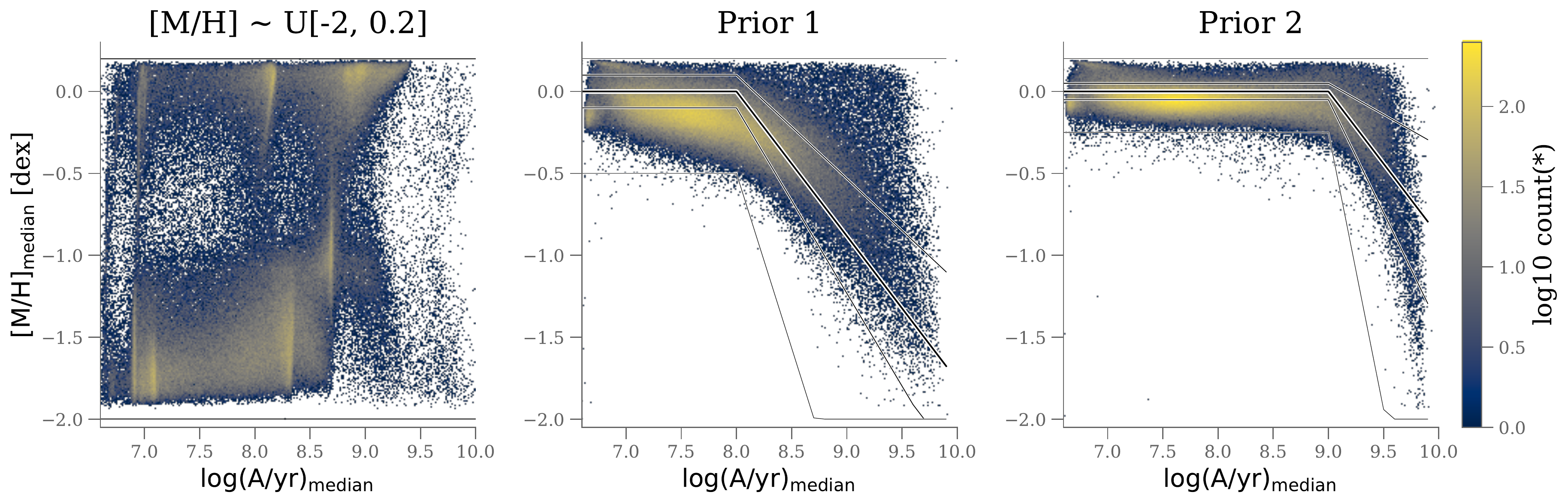}
       \end{center}
      \caption{
        Distribution of metallicity estimates as function of age compared to its prior. from left to right, each panel shows the posterior metallicity for random subset of stars in the direction of the Orion star formation using a uniform and age dependent prior, respectively. The median estimates are indicated by the point density, while the prior is shown by the solid lines (mean, 1, and 5 sigma intervals). Note that the prior rejects anything outside the model range (below $-2$ and above $0.2$ dex). The most right panel corresponds to our prior.
      }
      \label{fig:metallicity-prior-dominate}
\end{figure*}

As we mentioned before, it is not a surprise that we cannot constrain metallicity from out data.
The width of the Gaia passbands makes the photometric values barely changing from the variations of the optical metal lines.
Hence,  [Fe/H] becomes prior dominated.
Figure \ref{fig:metallicity-prior-dominate} illustrates the prior dependent estimates of metallicity [Fe/H].
If one would adopt a uniform prior (left panel), the impact would be substantial as the degeneracies create multi-modal solutions to match young stars' photometry. In regions like Orion on this figure, we can affirm the metallicity estimates are not plausible.

As a result, we adopted an age dependent prior (right panel) that allows some metallicity range at older ages as one would expect chemical enrichment to produce a similar trend to first order. We did not find these values to be scientifically exploitable individually. Our tests did not suggest we could see population variations within the Galaxy. As a result, we marginalized over them when reporting our estimates. This allows us to produce more meaninful uncertainties (see \autoref{sec:uncertainties}).

To improve upon the determination of metallicity, one could include narrower photometric bands (e.g., SDSS, PS1). This would lead to more crossmatch potential issues and potentially more tensions between models and data. \gdr{3} will contain the BP/RP spectra, i.e., the dispersed light corresponding to the optical coverage of Gaia. If one thinks of these spectra as series of narrow photometric bands, the metallicity issue should be resolved without these additional limitations.\footnote{The main challenge will become modeling the BP/RP spectra.}

\subsection{Photometric Jitter}

\begin{figure*}
    \begin{center}
        \includegraphics[width=2\columnwidth, clip, trim=0 0 0 1cm]{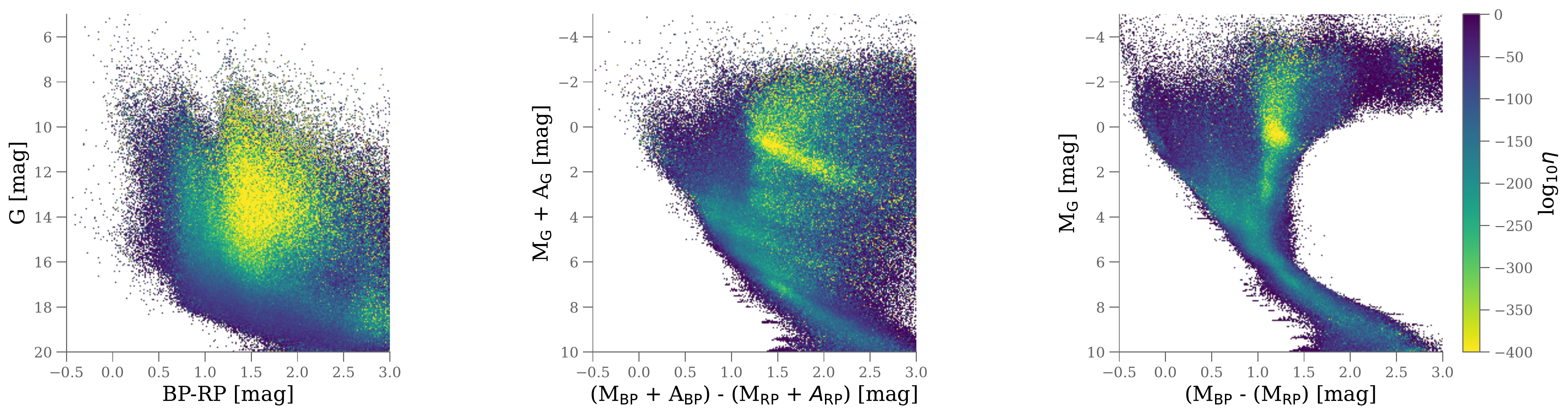}
       \end{center}
      \caption{
        Distribution of the photometric jitter $\eta$ for the whole \ntotcat\ stars, in the observed, distance corrected and distance-extinction corrected color magnitude diagrams, from left to right, respectively. Colorscale is identical for all three panels. Large values of the jitter indicates where models and data statistically deviate, i.e. mostly in rapid evolution phases. Note  the binary sequence parallel to the main sequence requires some substancial jitter values.
      }
      \label{fig:cmd-jitter}
\end{figure*}

We introduced a photometric Jitter $\eta$ in our model. This term acts as a random uncertainty in the photometry that we estimate per source.
It aims at capturing (random) mismatches between our models and data, which could be the result of a variety of reasons.
For example, photometric calibration errors could be large (e.g., $0.1\%$), or the data could be affected by crossmatch errors (especially in dense regions).
Fig.~\ref{fig:cmd-jitter} summarizes the jitter behavior. The left panel shows the distribution of the jitter in the observed Gaia color-magnitude diagram. One can immediatly see the prominence of the giants with small jitter values (yellow in this diagram). The other panels show the distributions after correcting the y-axis for distance and further dust extinction (middle and right panels, respectively).
The jitter becomes substantial for stars on short evolution stages, for instance OB stars of AGBs. One can also note that binaries (middle panel) also disagree with our models and thus force the jitter values to be larger than the single stars on the main sequence.

\section{Catalog}

\subsection{Content}\label{sec:catalog}
The catalog includes \ntotcat\ sources from \gdr{2} that have a parallax and all G, BP, RP, J, H, Ks, W1, and W2 photometric measurements.
Per star, we provide for each output quantity \texttt{X} the
\begin{itemize}
\item \texttt{X}: the input (recalibrated) quantity
\item \texttt{X\_sigma}: the input (recalibrated) uncertainty
\item  \texttt{X\_best}: the value of the best posterior sample from the MCMC chains
\item  \texttt{X\_min}, \texttt{X\_max}: min and max values of the MCMC samples
\item  \texttt{X\_p16}, \texttt{p50}, \texttt{p84}: 16, 50, 84th percentiles from the MCMC samples
\end{itemize}
the set \{\texttt{X\_best}\} represents the best model/prediction set (preserving physical information of the model).
In addition, we report the input features used during the fit that were “recalibrated” using the GDR2 prescriptions (see Table \ref{tab:calibtable}), i.e, photometry and parallaxes.
The catalog contains the following quantities
\begin{itemize}
\item \texttt{source\_id}: Gaia DR2 identifier
\item \texttt{parallax}: \texttt{parallax\_sigma}: recalibrated parallax and uncertainty values
\item \texttt{J}: 2MASS J photometry [vegamag],
\item \texttt{H}: 2MASS H band photometry [vegamag],
\item \texttt{Ks}: 2MASS Ks photometry [vegamag],
\item \texttt{BP}: Gaia BP magnitude (bright or faint) [vegamag]
\item \texttt{G}: Gaia G magnitude [vegamag]
\item \texttt{RP}: Gaia RP magnitude [vegamag]
\item \texttt{W1}: AllWise W1 magnitude [vegamag]
\item \texttt{W2}: AllWise W2 magnitude [vegamag]
\item \texttt{A0}: extinction parameter [mag]
\item \texttt{R0}: average dust grain size extinction parameter [unitless]
\item \texttt{A\_G}: extinction in the Gaia G-band [mag]
\item \texttt{A\_BP}: extinction in the Gaia BP-band [mag]
\item \texttt{A\_RP}: extinction in the Gaia RP-band [mag]
\item \texttt{dmod}: distance modulus [mag]
\item \texttt{lnlike}: log likelihood [unitless]
\item \texttt{lnp}: log posterior [unitless]
\item \texttt{log10jitter}: log photometric likelihood jitter common to all bands [log10 mag]
\item \texttt{logA}: log10(age/yr)
\item \texttt{logL}: log10(Luminosity/L$_\odot$)
\item \texttt{logM}: log10(mass/M$_\odot$)
\item \texttt{logT}: log10(Teff/K)
\item \texttt{logg}: log10(gravity/cgs)
\end{itemize}

Note that from \gdr{2} to \gdr{3}, one should recall that 3\% of the sources {\tt source\_id} identifier were updated \citep{Fabricius2020}.
Thus, one should use the {\tt source\_id} crossmatch table {\tt dr2\_neighbourhood} provided with \gdr{3} to find the best match before any source-by-source comparisons between the two releases.

We have not filtered out any results from our catalog. Sources with spurious parallaxes remain. One must proceed to any filtering with care; any rules are most unlikely to generalize. Instead, selections need to adapt to specific use-cases and locations in the Galaxy.

\subsection{Use-cases}

Some example use cases are as follows.

\begin{enumerate}
\item Look-up of distance (or distance modulus) for particular sources of interest using their {\tt source\_id} or other identifier matched to this. The Gaia data releases include a crossmatch to many existing catalogs. Positional crossmatch can also be done on the data site or using TAP uploads and at other sites that host our catalog.

\item Identification of sources within some astrophysical parameter ranges. One should use the confidence intervals to find all sources of interest. For instance, \citet{Poggio2021} select upper main sequence stars from their apparent colors; one could switch to \teff, or the absolute magnitude predictions from our catalog.

\item Construction of color-absolute-magnitude diagrams (CAMDs).
One of the reasons why we provide quantiles on distance modulus and the individual photometric band predictions and their extinction values. (This would not be possible if we reported the mean or mode, for example.)

\item For constructing the three-dimensional spatial distribution of stars in some region of space. It may also assist selections of candidates in targeted follow-up surveys.

\item For constructing the three-dimensional spatial properties of the ISM.
Using our extinctions and distances, \citet[in press]{Dharmawardena2021}  inferred the individual structure of the Orion, Taurus, Perseus, and Cygnus X star-forming regions and found the coherent ISM filaments that may link the Taurus and Perseus regions.

\item Linking \Rzero's spatial variations to local stellar properties to understand ISM processing cycles \citep[e.g.,][]{Chastenet2017}.

\item As a baseline for comparison of APs or absolute magnitude estimates inferred by other means.

\end{enumerate}

\subsection{Data Access}\label{sec:access}

Our catalog is available from the German Astrophysical Virtual Observatory (GAVO)\footnote{GAVO:
\url{http://dc.g-vo.org/tableinfo/gdr2ap.main}} where one can query it via ADQL and TAP.
This server also hosts a reduced version of the main Gaia \gdr{2} and \gdr{3}catalogs. Typical queries are likely to involve a join of one of these two catalogs.
A bulk download for the catalog is also available at the URL given above. Our catalog will also become available soon in the Gaia Archive \footnote{Gaia Archive \url{https://gea.esac.esa.int/archive/}} and CDS VizieR and their partner data centers.

\subsection{Limitations \& Discussions}\label{sec:discussion}

Users should keep in mind its assumptions and limitations when using our catalog.

\begin{itemize}
\item We summarize the 6-dimensional posterior distributions using only quantile numbers (computed on marginal 1D distributions). The \texttt{\_best} estimates only give one point that was randomly sampled and turned out to obtain the best posterior value of the MCMC chains. These summary statistics cannot capture the full complexity of these distributions. One should not ignore the confidence intervals.

\item Most sources in \gdr{2} have substantial fractional parallax uncertainties. Hence, the photometric data often dominate the inference of our distances and APs. However, the parallax remains generally sufficient to limit the dwarf vs. giant degeneracies.

\item The poorer the data, the more our prior dominates our estimates. Our prior is \textit{not} a sophisticated model of the Galaxy that includes 3D extinction. One should expect significant differences with other AP catalogs when prior dominates. However, in reality, if the true stellar population, extinction, or reddening distributions are very different from Galactic models, our catalog (and disagreements with other studies) may partially hint at these deviations.

\item We implicitly assumed that all sources are single stars in the Galaxy.
Our estimates will be incorrect for any non-single star (binaries, extended sources, extragalactic).

\item We assumed all crossmatch to produce correct results. Matching sources becomes a complex challenge in some cases (e.g., dense regions, high extinction).  Wrong associations between surveys result in incorrect photometric data, leading to ``bad fits''. We introduced $\eta$ to capture some of these issues. In combination with \texttt{lnlike} in our catalog, one could assess how the model predictions were matching the sources' photometric information.

\item By design, we infer properties for each source independently. If a set of stars is known to be in a cluster, they have a similar distance, extinction, chemical patterns, and age. It constitutes a prior that one should exploit to infer the properties of the individual stars more accurately than what we have done here.
\end{itemize}

\section{Summary}\label{sec:summary}

We have produced a catalog of distance moduli, astrophysical and dust extinction parameters for \ntotcat\ stars using the photometry from Gaia, 2MASS, and AllWise and the Gaia parallaxes. These estimates, and their uncertainties, can also be used as estimates of the distances. We provide additional photometric estimates and their dust {extinction} values.

Our catalog increases the availability of APs in the literature while offering results based on different assumptions than previous works. The latter helped to validate our results.

In addition, we provide one of the first extensive catalogs of average {extinction} per color-excess unit (\Rzero) derived uniformly across the whole sky.

We used external data, so-called ``hybrid catalogs'', to leverage multi-wavelength current information from independent data sources. However, combining heterogeneous data inevitably leads to inconsistencies, biases, and complex selection functions.

With \gdr{3}, DPAC will publish APs based upon analyzing the BP/RP spectra, i.e., Gaia data only with detailed optical information according to the \href{https://www.cosmos.esa.int/web/gaia/release}{Gaia data release Scenario}. Such a product will eventually open the possibility to further ``hybrid'' analysis. Gaia DR3 represents a critical first step to anchor all current and future spectroscopic surveys to a common ground and provide us with the most comprehensive view of our Galaxy.

\begin{acknowledgements}
We thank our referee for their constructive and valuable comments.
We also thank H.-W. Rix and D.~W. Hogg for the fruitful discussions during this project. We also thank C. Subiran for providing us with compiled catalogs from the literature. MF thanks K. Gordon for sharing his expertise on dust properties.
\\
This work was funded in part by the DLR (German space agency) via grant 50 QG
1403.
It has made use of data from the European Space Agency (ESA) mission Gaia (\url{http://www.cosmos.esa.int/gaia}), processed by the Gaia Data Processing and Analysis Consortium (DPAC, \url{http://www.cosmos.esa.int/web/gaia/dpac/consortium}). Funding for the DPAC has been provided by national institutions, in particular the institutions participating in the Gaia Multilateral Agreement.
This publication uses data products from the Two Micron All Sky Survey, a joint project of the University of Massachusetts and the Infrared Processing and Analysis Center/California Institute of Technology, funded by the National Aeronautics and Space Administration and the National Science Foundation.
This publication uses data products from the Wide-field Infrared Survey Explorer, a joint project of the University of California, Los Angeles, and the Jet Propulsion Laboratory/California Institute of Technology funded by the National Aeronautics and Space Administration.
\\
This research made use of
matplotlib \citep{Hunter:2007},
NumPy \citep{harris2020array},
the IPython package \citep{PER-GRA:2007},
Vaex \citep{Breddels2018},
TOPCAT \citep{Taylor2005},
QuantStack xtensor (\url{http://xtensor.readthedocs.io/}).
\end{acknowledgements}



\bibliographystyle{aa} 
\bibliography{bibliography} 

%

\begin{appendix}

\section{Validation target sample}
\label{appendix:VST-definition}

\noindent
This appendix briefly describes how we compile a ``reference catalog'' from various literature resources. First, we present the general quality requirements from the literature values that any star needs to satisfy to enter our reference catalog. Afterward, we go over the various literature catalogs and describe their additional specific selection criteria. Table~\ref{table:composition-of-master-catalog} provides a summary of the diverse literature catalog contributions to our reference sample.

\begin{table}
\caption{Contributions to reference catalog from various literature catalogs.}
\label{table:composition-of-master-catalog}
\begin{center}
\begin{footnotesize}
\begin{tabular}{lrr}
catalog & all stars & selected stars \\
\hline
\apogee & 277\,371 & 69\,572 \\
\ges & 82\,123 & 36\,159 \\
\galah & 342\,682 & 223\,089 \\
\lamost & 4\,540\,986 & 3\,251\,173 \\
\rave & 520\,701 & 123\,988 \\
\sdss & 430\,164 & 309\,934 \\
\kicA & 6\,569\,685 & 1\,704\,630 \\
\kicB & 197\,096 & 185\,321 \\
\cu6 & 1\,930\,105 & 368\,153 \\
\redclump & 19\,937 & 19\,937 \\
\obRamirez & 72 & 72 \\
\obSimon & 382 & 382 \\
\hline
total &    & \ntotvstcat \\
unique Gaia sources &    & 4\,751\,603
\end{tabular}
\end{footnotesize}
\end{center}
\end{table}

\subsection{General quality requirements for all catalogs}
\label{ssect:general-quality-requirements}

For any star with literature properties, the following general quality requirements must be satisfied to admit that star into our reference catalog:
\begin{itemize}
\item the star must have estimated values for \textit{all} of the three stellar parameters \teff, \logg, [M/H].
\item the star must have uncertainties estimated for \textit{all} of the three stellar parameters \teff, \logg, [M/H].
\end{itemize}

\subsection{Catalog-specific selection criteria}

\paragraph{\apogee} - \citet{ApogeeDR14}

we take the data file \texttt{allStar-l31c.2.fits}. It contains 277\,371 stars with ASPCAP parameter estimates. We apply the following quality cuts:
\begin{itemize}
\item \texttt{ASPCAPFLAG} is not $2^{23}$, which excludes ``bad overall for star: set if any of \texttt{TEFF}, \texttt{LOGG}, \texttt{CHI2}, \texttt{COLORTE}, \texttt{ROTATION}, \texttt{SN error} are set, or any parameter is near grid edge''.
\item \texttt{COMMISS} equals $0$, which excludes \apogee-1 commissioning data.
\item \texttt{\apogee\_TARGET1} is not $2^{11}+2^{12}+2^{13}$, which excludes short/intermediate/long cohort target stars from \apogee-1, i.e., commissioning data again.
\item Signal-to-noise ratio larger than 50.
\end{itemize}
This filtering reduces the number of selected stars down to 69\,572, which contains several hundred duplicated entries, one can identify via their \apogee~IDs.

\paragraph{\ges} -- \citet{Gilmore2012}

Starting from the internal DR5 with 82\,123 stars, we apply the general quality requirements without any additional cuts. The final sample contains 36\,159 stars.

\paragraph{\galah} -- \citet{Buder2018}

\galah\ published stellar parameters for 342\,682 stars, including uncertainty estimates.  We apply the following quality cuts (private communication with S. Buder):
\begin{itemize}
\item \texttt{flag\_cannon} equals $0$, which excludes spectra that are unusual (e.g., binaries) or had a problematic reduction, or outside the model validity (e.g., cool stars).
\item \texttt{snr\_c2} larger than $25$ (in the green band; the official \galah end-of-survey goal is 50).
\item $\teff<7\,000$K, limitation from the \textsc{Cannon}'s empirical training sample under-representing hotter stars. (Cool stars are caught through the flag already.)
\end{itemize}
The general and additional quality cuts leave 223\,089 stars.

\paragraph{\lamost} -- \citet{2011RAA....11..924W,2014IAUS..306..340W}

The \lamost\footnote{\url{http://dr4.\lamost.org/catalog}} contains parameter estimates for 4\,540\,986 AFGK stars.
We apply the general quality requirements and the following additional cuts (private communication with C. Liu):
\begin{itemize}
\item Signal-to-noise ratio of g filter \texttt{snrg} larger than 20.
\item Estimated temperature in range 4\,000K$<\teff<$7\,500K.
\end{itemize}
We obtain with 3\,251\,173 stars.

\paragraph{\rave} -- \citet{2017AJ....153...75K,2013AJ....146..134K}

The \rave~contains parameter estimates for 520\,701 stars. We apply the additional quality cuts:
\begin{itemize}
\item Stellar classification 1 equals \texttt{d}, \texttt{g}, \texttt{h}, \texttt{n} or \texttt{o}.
\item Stellar classification 2 equals \texttt{d}, \texttt{g}, \texttt{h}, \texttt{n} or \texttt{o}.
\item Stellar classification 3 equals \texttt{d}, \texttt{g}, \texttt{h}, \texttt{n} or \texttt{o}.
\item \texttt{algo\_conv} equals 0.
\item Error of Heliocentric radial velocity is less than $10$\,km/s.
\item Signal-to-noise ratio in $K$-band is larger than 50.
\item $4000<\teff<7750$
\item $\log g>1$
\item Removal of multiple entries, keeping the one with highest signal-to-noise ratio in $K$-band.
\end{itemize}
These cuts leave 123\,988 stars, for which we use the calibrated parameter estimates only.

\subsubsection{\sdss~stars} -- \citet{ApogeeDR14}

For \sdss, we queried their CAS server with the following:
\footnote{\url{http://cas.\sdss.org/dr14/en/tools/search/sql.aspx}}

\begin{verbatim}
SELECT
   p.objid,
   p.ra, p.dec,
   p.u, p.g, p.r, p.i, p.z,
   p.run, p.rerun, p.camcol, p.field,
   s.specobjid,
   s.plate, s.mjd, s.fiberid,
   s.TEFFADOP,s.TEFFADOPUNC,
   s.TEFFSPEC,s.TEFFSPECUNC,
   s.LOGGADOP,s.LOGGADOPUNC,
   s.LOGGSPEC,s.LOGGSPECUNC,
   s.FEHADOP,s.FEHADOPUNC,
   s.FEHSPEC,s.FEHSPECUNC,
   s.FLAG,s.SNR,s.QA
FROM PhotoObj AS p
   JOIN sppParams AS s ON s.bestobjid = p.objid
WHERE
   p.g BETWEEN 0 AND 21
   AND (s.TEFFADOP>0.0 OR s.TEFFSPEC>0.0)
\end{verbatim}
This request produced 430\,164 stars. We use the ``spectroscopic'' parameters (not the ``adopted''). In addition to the general quality requirements, we also selected the \texttt{FLAG} equals to \texttt{nnnnn}, i.e., ``normal''. This resulted in 309\,934 selected stars.

\paragraph{Raw \& Revised Kepler Input Catalog} -- \citet{2010AN....331..981M} \& \citet{Mathur2017}

For the \kicA, we downloaded the online data file \texttt{kic\_ct\_join\_12142009.txt}\footnote{\url{https://archive.stsci.edu/pub/kepler/catalogs/kic_ct_join_12142009.txt.gz}}. It contains 6\,569\,685 stars.
We caution that this catalog has limitations \citep{2010AN....331..981M} and thus one should use it with caution. Unfortunately, this does not contain any uncertainty estimates on stellar parameters, such that we cannot apply the general quality requirements. Nonetheless, requiring all three stellar parameters (\teff, \logg,, [M/H]) leads us to 1\,704\,630 stars in this sample.

\citet{Mathur2017} provides revised stellar properties for 197\,096 Kepler targets. After applying our general quality requirements, we obtain 185\,321 stars.
We replaced any duplicated entry with the raw input catalog with the parameters of this catalog.

\paragraph{\redclump} -- \citet{Bovy2014}

We downloaded the \apogee~red-clump catalog\footnote{\url{data.\sdss3.org/sas/dr12/\apogee/vac/\apogee-rc/cat/\apogee-rc-DR12.fits}} which contains 19\,937 red-clump stars satisfying our general quality requirements.

\paragraph{OB stars} -- \citet{RamirezAgudelo2017} \& \citet{SimonDiaz2017}

\citet{RamirezAgudelo2017} publish effective temperatures, $\log g$, radius, mass, and bolometric luminosities for 72 OB stars. Unfortunately, they do not publish estimates of [M/H].
\citet{SimonDiaz2017} publish effective temperatures and bolometric luminosities for 382 OB stars. Unfortunately, they do not publish estimates of $\log g$ or [M/H] nor their uncertainty estimates.

Nevertheless, since OB stars with parameters are scarce, we still admit both sets of 72 and 382 OB stars into our reference catalog.

\paragraph{Gaia \cu6} -- \citet[private communication]{LL:CS-011}

We adopt the Gaia DPAC \cu6 compiled to validate Gaia RVS spectra and CU8 stellar parameter estimates. In its latest version (\texttt{V240114}), it contains 1\,930\,105 stars. We apply the general quality cuts, but we do not insist on uncertainty estimates on $\log g$ or [Fe/H]. This selection leaves us with 368\,153 stars.

\subsection{Crossmatch to \gdr{2}}

Given the reference catalog of \ntotvstcat~stars compiled from the literature, we crossmatch the coordinates to \gdr{2}. Further restrictions for the crossmatch process are:
\begin{itemize}
\item an apparent magnitude $G\leq 20$\,mag.
\item at least $2$ transits of the sources observed by Gaia,
\item a maximum crossmatch distance of $3$\,arcsec,
\item and if the target source has an expected $G$ magnitude, we only consider sources within $|\Delta G|<1$\,mag.
\end{itemize}
Of the total number of \ntotvstcat~targets from the reference catalog, we obtain found 5\,080\,846 matches in \gdr{2}. Still, duplicate Gaia source IDs originate from the same star occurring in more than one literature catalog. The final reference catalog contains 4\,751\,603 unique Gaia source IDs.

\section{Additional figures}

In this section, we show the following figures:
\begin{itemize}
  \item Fig.~\ref{fig:cmdcheck} is the equivalent of Fig.~\ref{fig:vstcmdcheck} for the \ntotcat\ stars of the whole catalog.
  \item Fig.~\ref{fig:cmddustcheck} is the equivalent of Fig.~\ref{fig:vstcmddustcheck} for the \ntotcat\ stars of the whole catalog.
  \item Fig.~\ref{fig:cmdcheck_residuals} presents the residuals between the observed and inferred color-magnitude diagrams (top panels of Fig.~\ref{fig:cmdcheck}) for the \ntotcat\ stars of the whole catalog.
  \item Fig.~\ref{fig:mw_a0_r0_maps_zooms} and Fig.~\ref{fig:mw_a0_r0_maps_zooms_2} show zoomed in versions of the regions highlighted in Fig.~\ref{fig:mw_a0_r0_maps}.
\end{itemize}

\begin{figure*}
    \begin{center}
        \includegraphics[width=1.8\columnwidth, clip]{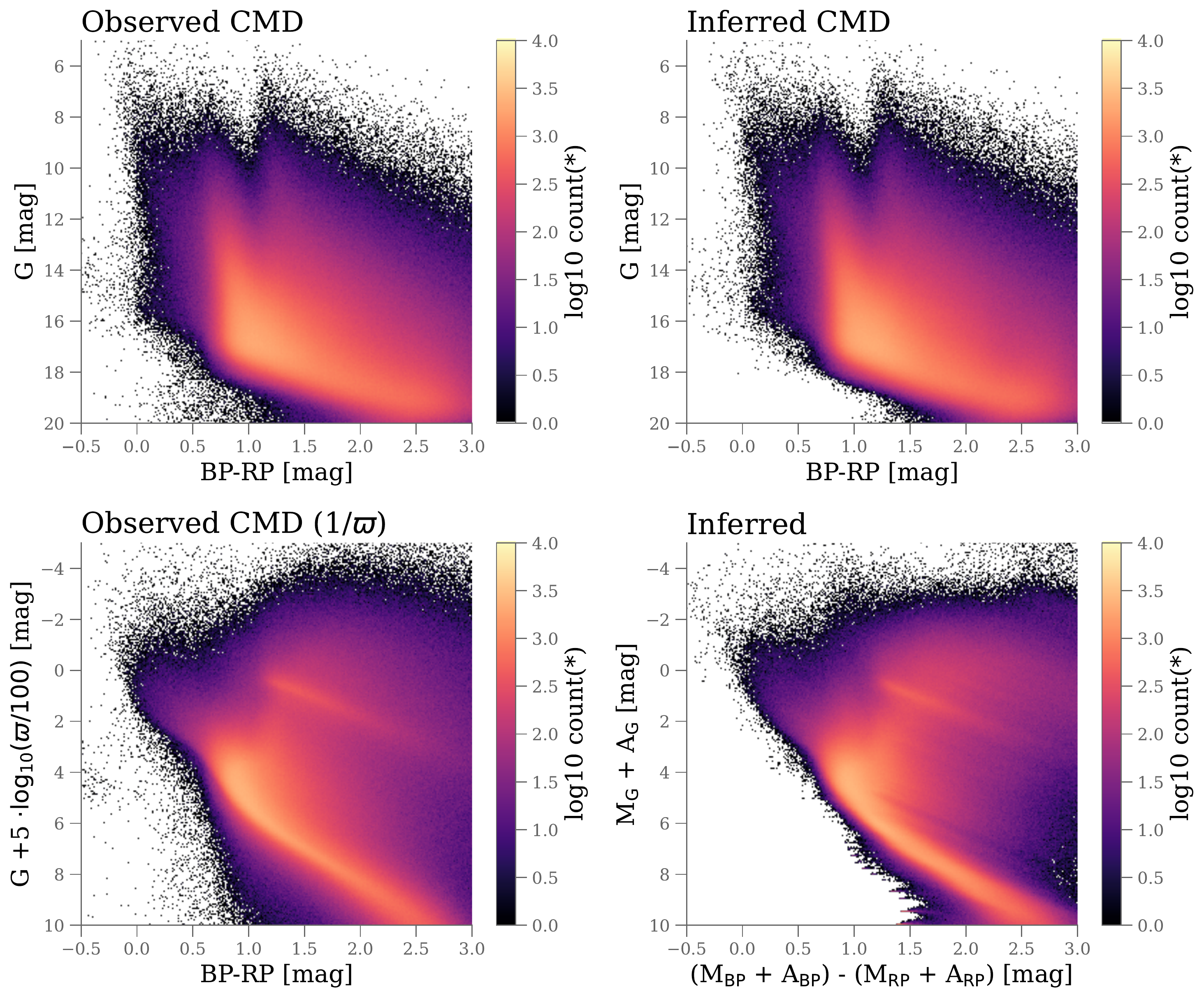}
      \end{center}
      \caption{Overview our analysis procedure on the whole catalog containing \ntotcat\ stars. Top panels present observed color-magnitude diagrams where left and right panels are the input data and their median predictions, respectively. The lower panels show the inverse parallax distance corrected CMD and the distance corrected obtained from the AP estimates. The quantity on the y-axis of these two panels would be identical in absence of parallax noise.
      }
      \label{fig:cmdcheck}
\end{figure*}

\begin{figure*}
  \begin{center}
      \includegraphics[width=1.8\columnwidth, clip]{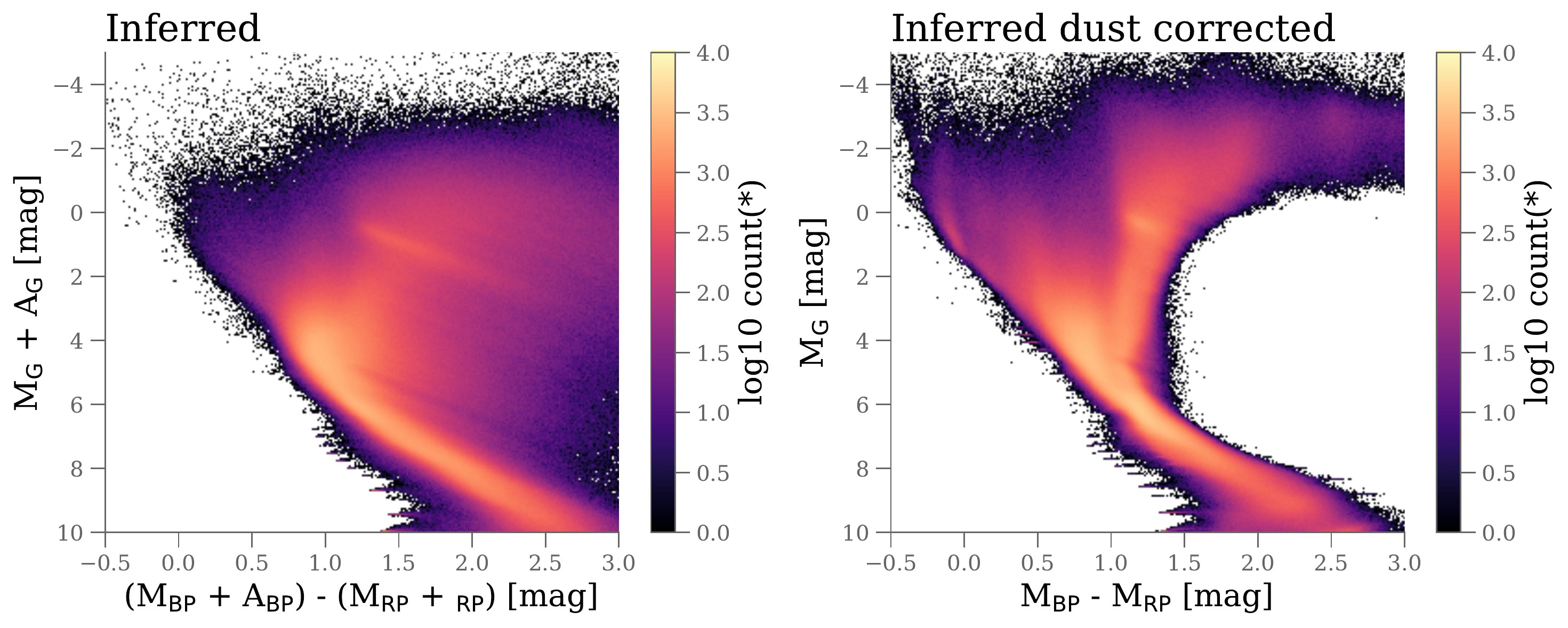}
    \end{center}
    \caption{
      Inferred color absolute magnitude diagram (CAMD) before (left) and after (right) accounting for the dust extinction for the entire catalog.
    }
    \label{fig:cmddustcheck}
\end{figure*}

\begin{figure*}
  \begin{center}
      \includegraphics[width=\columnwidth, clip, trim=30cm 0 1.1cm 1.1cm]{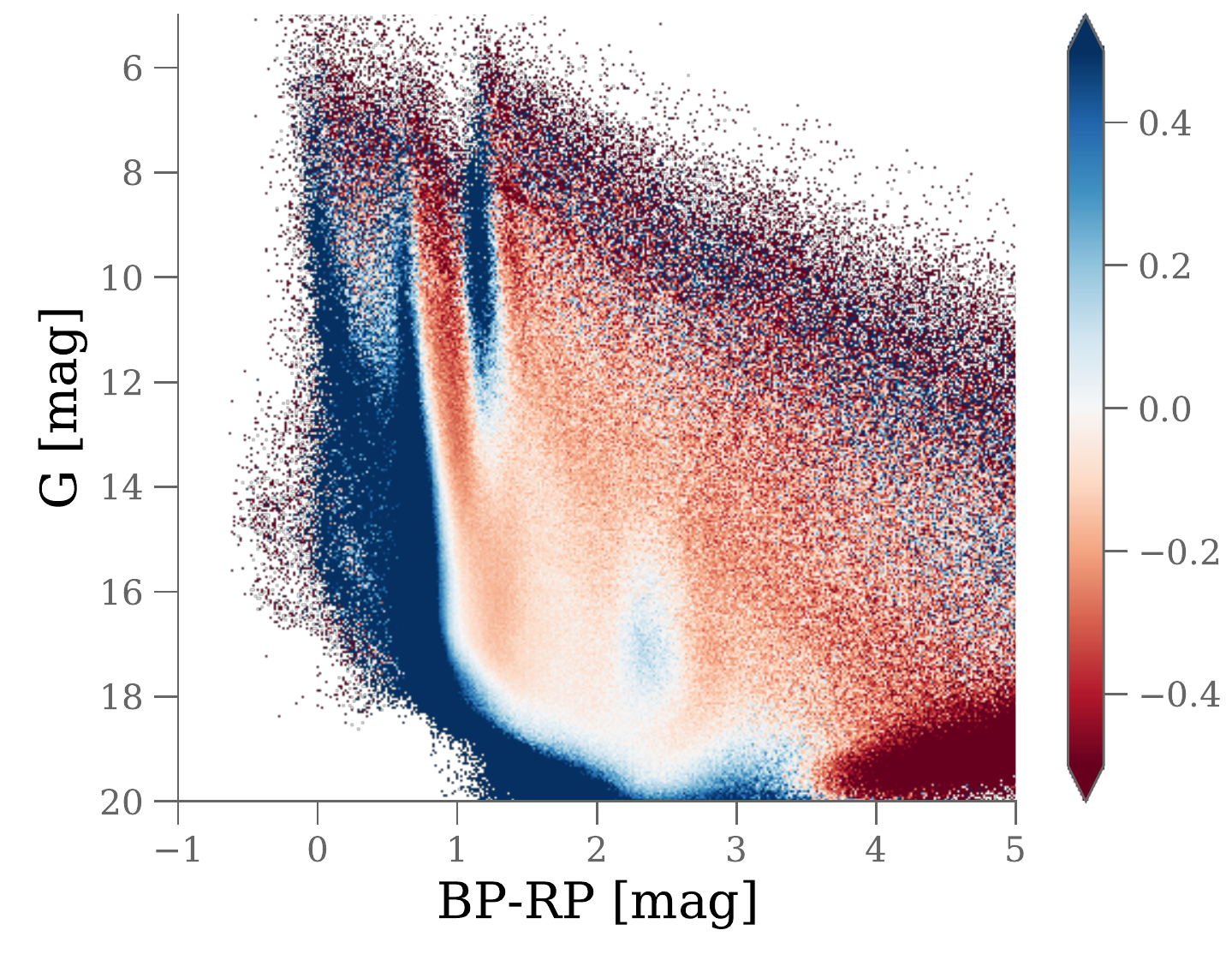}
    \end{center}
    \caption{
      Residuals between the observed and inferred color magnitude diagrams (CAMD) for the \ntotcat\ stars of the whole catalog (top panels of Fig.\ref{fig:cmdcheck}). The residuals are calculated on a 512x512 binning (0.01 x 0.03 mag) scheme within the axis ranges and normalized to the predicted counts (i.e., (obs-pred / pred)): red and blue colors indicate overestimated and underestimated counts, respectively. Systematics are discussed in Sect.\,\ref{sec:validation}.
      Stars bluer than what the stellar evolution model predict pile up in the residuals at BP-RP $\sim$ 1\,mag.
    }
    \label{fig:cmdcheck_residuals}
\end{figure*}

\begin{figure*}
  \begin{center}
    \includegraphics[width=1.8\columnwidth, clip]{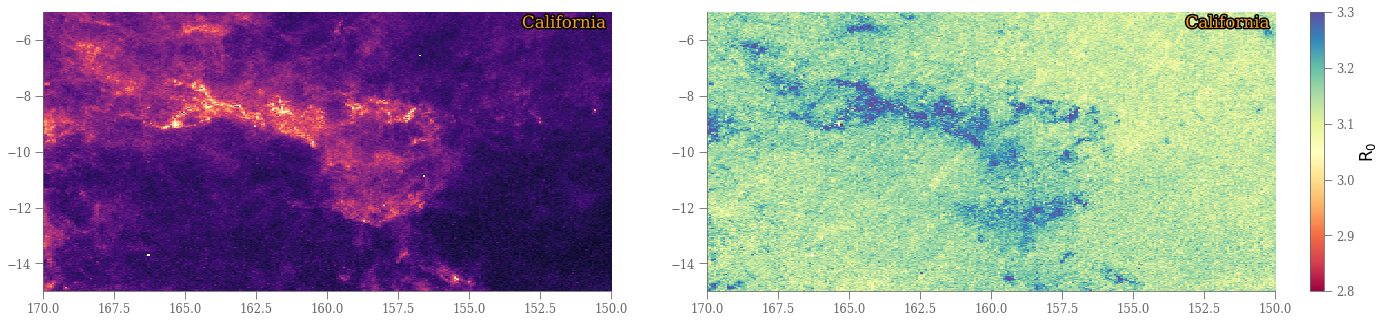}
    \includegraphics[width=1.8\columnwidth, clip]{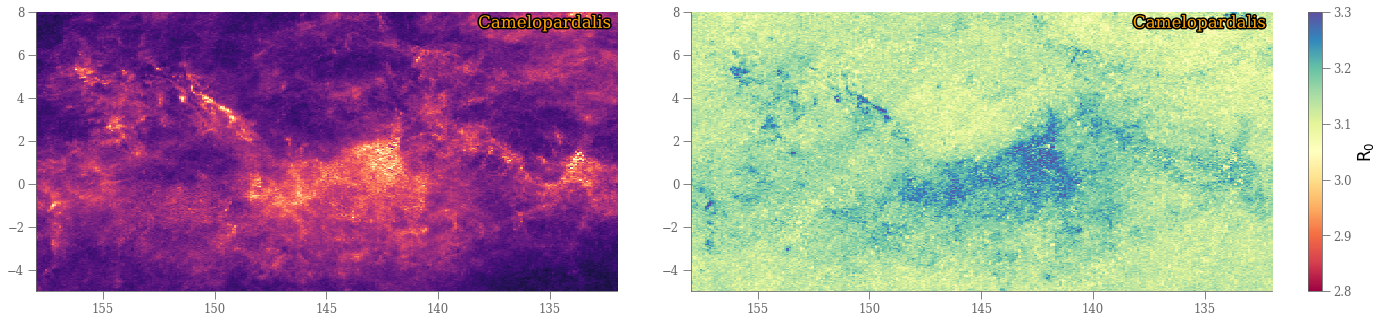}
    \includegraphics[width=1.8\columnwidth, clip]{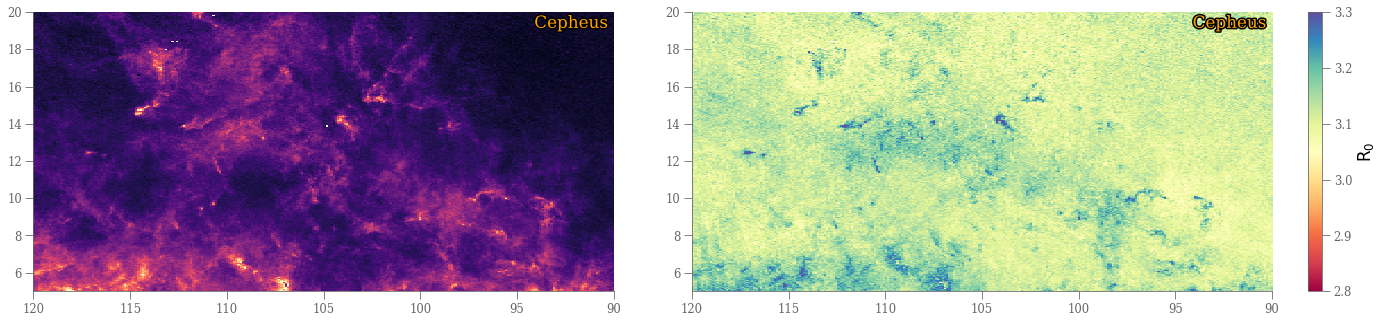}
    \includegraphics[width=1.8\columnwidth, clip]{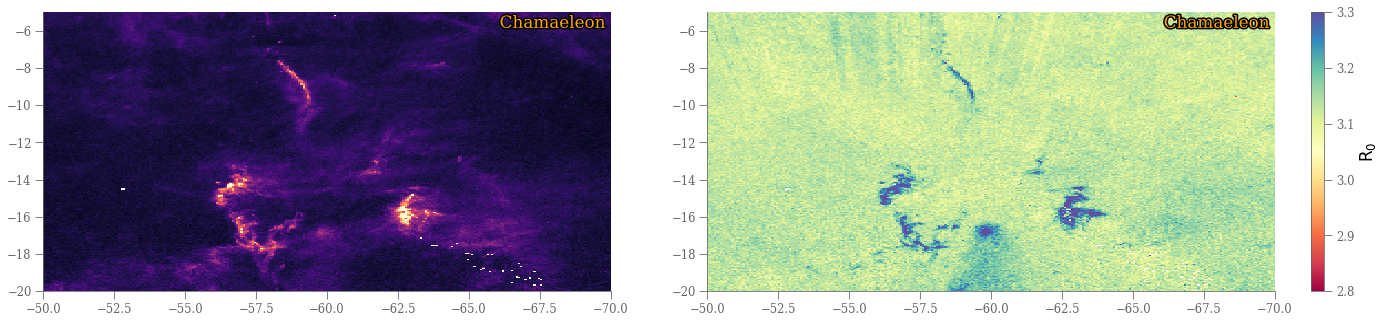}
    \includegraphics[width=1.8\columnwidth, clip]{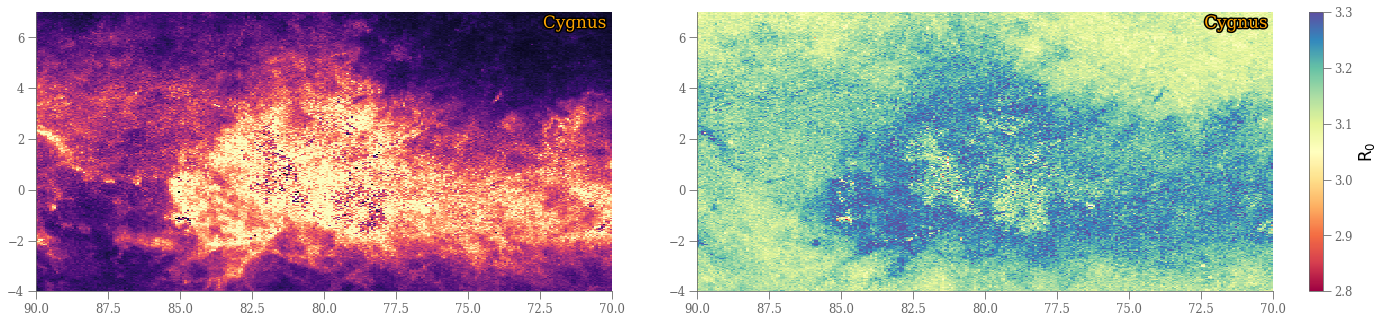}
    \includegraphics[width=1.8\columnwidth, clip]{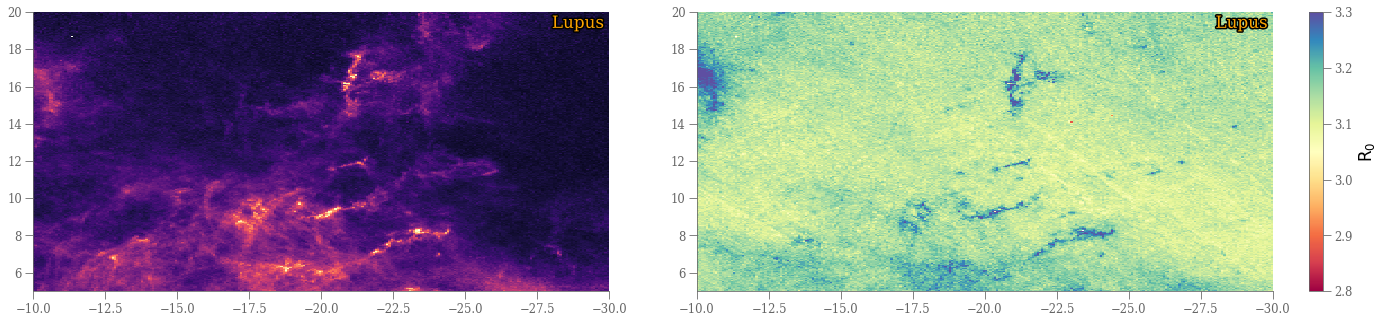}
    \end{center}
    \caption{
    Sky distribution in Galactic coordinates (averaged over all distances) of the dust extinction parameters \Azero (left) and \Rzero (right) of the regions highlighted in Fig.~\ref{fig:mw_a0_r0_maps}.
    The region's names are indicated in the top-right corner of each panel. Continues in Fig.~\ref{fig:mw_a0_r0_maps_zooms_2}
    }
    \label{fig:mw_a0_r0_maps_zooms}
\end{figure*}
\begin{figure*}
  \begin{center}
    \includegraphics[width=1.8\columnwidth, clip]{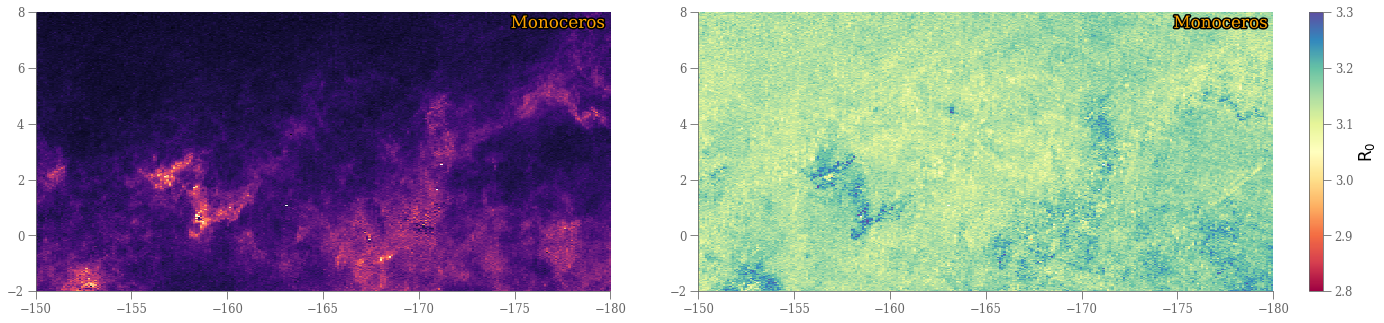}
    \includegraphics[width=1.8\columnwidth, clip]{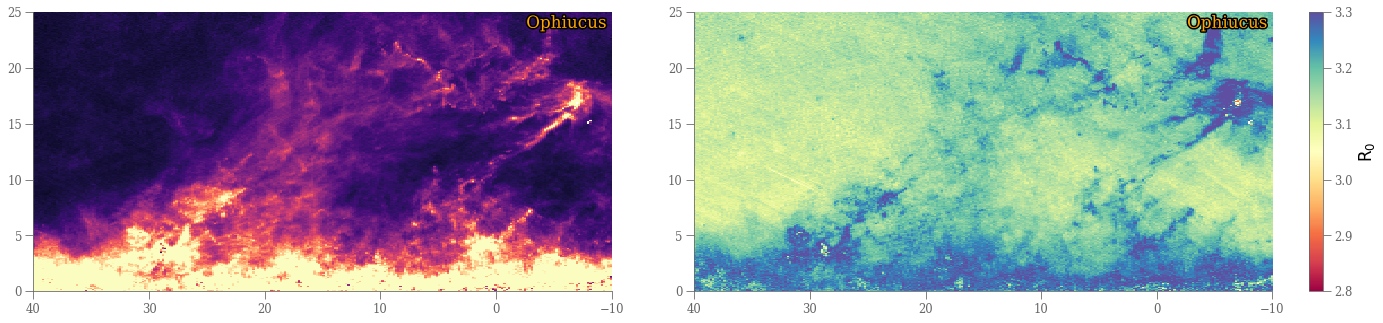}
    \includegraphics[width=1.8\columnwidth, clip]{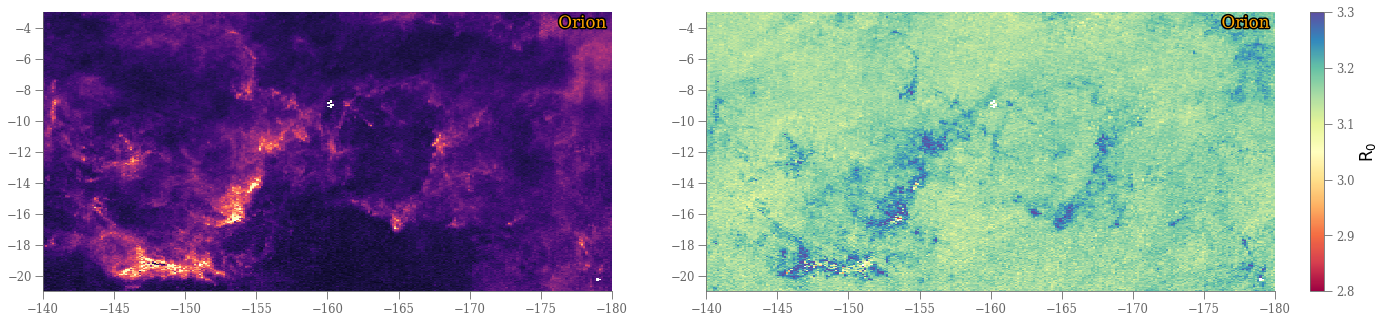}
    \includegraphics[width=1.8\columnwidth, clip]{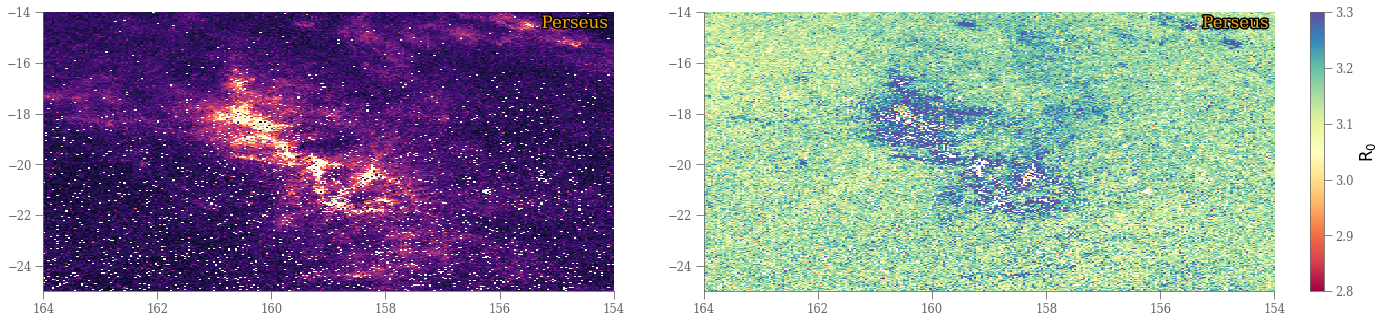}
    \includegraphics[width=1.8\columnwidth, clip]{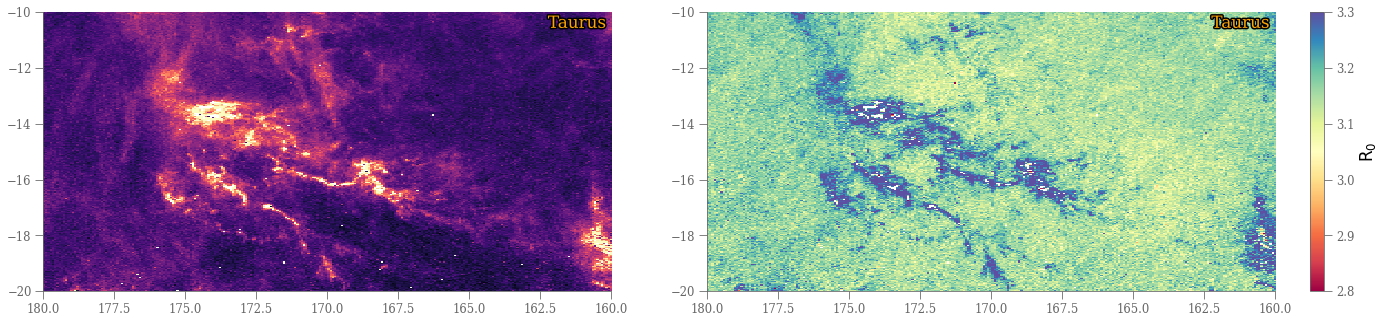}
    \includegraphics[width=1.8\columnwidth, clip]{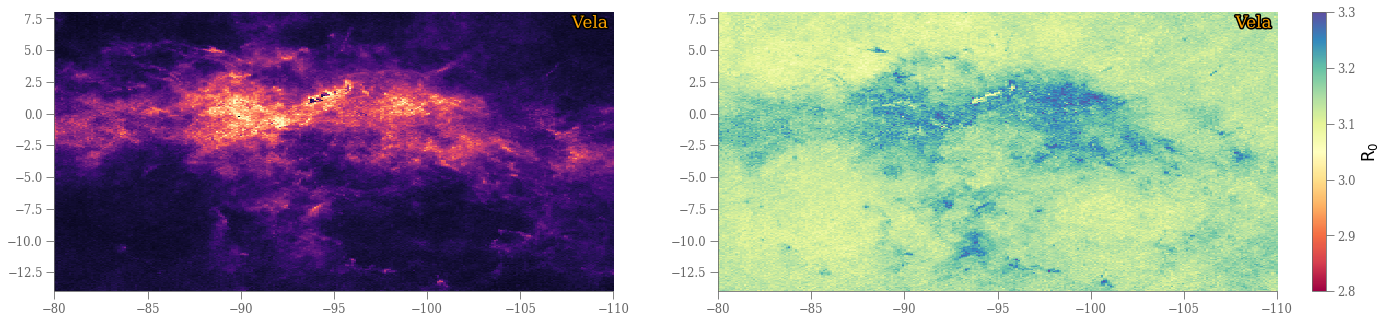}
    \end{center}
    \caption{
    Continuation of Fig.~\ref{fig:mw_a0_r0_maps_zooms}.
    }
    \label{fig:mw_a0_r0_maps_zooms_2}
\end{figure*}

\end{appendix}

\end{document}